\documentclass[preprint]{aastex}

\begin{document}

\shorttitle{Dust Disks in Taurus-Auriga}

\shortauthors{Andrews \& Williams}

\title{Circumstellar Dust Disks in Taurus-Auriga: The Submillimeter Perspective}

\author{Sean M. Andrews \and Jonathan P. Williams}

\affil{Institute for Astronomy, University of Hawaii, 2680 Woodlawn Drive, Honolulu, HI 96822}
\email{andrews@ifa.hawaii.edu, jpw@ifa.hawaii.edu}

\begin{abstract}
We present a sensitive, multiwavelength submillimeter continuum survey of 153 
young stellar objects in the Taurus-Auriga star formation region.  The 
submillimeter detection rate is 61\% to a completeness limit of $\sim$10\,mJy 
(3-$\sigma$) at 850\,$\mu$m.  The inferred circumstellar disk masses are 
log-normally distributed with a mean mass of $\sim 5 \times 
10^{-3}$\,M$_{\odot}$ and a large dispersion (0.5\,dex).  Roughly one third of 
the submillimeter sources have disk masses larger than the minimal nebula from 
which the solar system formed.  The median disk to star mass ratio is 0.5\%.  
The empirical behavior of the submillimeter continuum is best described as 
$F_{\nu} \propto \nu^{2.0 \pm 0.5}$ between 350\,$\mu$m and 1.3\,mm, which we 
argue is due to the combined effects of the fraction of optically thick 
emission and a flatter frequency behavior of the opacity compared to the 
interstellar medium.  This latter effect could be due to a substantial 
population of large dust grains, which presumably would have grown through 
collisional agglomeration.  In this sample, the only stellar property that is 
correlated with the outer disk is the presence of a companion.  We find 
evidence for significant decreases in submillimeter flux densities, disk 
masses, and submillimeter continuum slopes along the canonical infrared 
spectral energy distribution evolution sequence for young stellar objects.  The 
fraction of objects detected in the submillimeter is essentially identical to 
the fraction with excess near-infrared emission, suggesting that dust in the 
inner and outer disk are removed nearly simultaneously.  
\end{abstract}

\section{Introduction}

The formation and early evolution of stars are intimately coupled to the 
properties of their accompanying circumstellar disks of gas and dust.  These 
disks also provide the material reservoirs for the assembly of planetary 
systems.  Angular momentum conservation dictates that a collapsing molecular 
cloud core with some initial rotation will result in both a central protostar 
and a flattened circumstellar disk \citep[e.g.,][]{terebey84}.  Indirect 
observations indicate that disks are essentially ubiquitous in young star 
clusters, while optical images in silhouette \citep{odell94} and millimeter 
spectral line confirmations of Keplerian rotation \citep[e.g.,][]{simon00} 
provide more direct evidence in specific cases.  Comparisons of infrared 
observations with physical models of young stellar objects (YSOs; here taken to 
mean a young star and its associated circumstellar material) have led to a 
sequence of evolutionary stages which occur before the start of the 
main-sequence \citep{lada84,adams86,adams87}.  In the Class I stage, an 
extended circumstellar envelope is rapidly dumping material onto a central 
protostar and a massive accretion disk.  After the supply of envelope material 
is dissipated, the YSO becomes a Class II object, with a disk that is actively 
accreting material onto a central, optically visible star.  In the final Class 
III stage, \emph{at least} the inner part of the circumstellar disk has been 
evacuated, although the dominant physical mechanism for this process remains in 
debate \citep[see][]{hollenbach00}.  The most interesting possibility, at least 
from a cosmogonical viewpoint, is that the gas and dust in the disk have 
agglomerated into larger objects in a developing planetary system.

Observations of the morphology of the broadband spectral energy distribution 
(SED) and various diagnostics of accretion can be used to trace the evolution 
of a YSO.  Longward of $\sim$1\,$\mu$m, the SED of a YSO is composed of a 
continuum of thermal spectra from the radially distributed circumstellar dust, 
modified by the radiative transfer properties of the grains.  Changes in the 
SED through the evolutionary sequence are indicative of the \emph{loss} of 
circumstellar components in the system; first the envelope and then the disk.  
The slope of the infrared SED is determined by the radial temperature 
distribution of the circumstellar dust \citep[e.g.,][]{adams87,bscg90}.  
Therefore, measurements of infrared colors provide a relatively simple 
observational constraint on the temperature structure of a disk.  However, more 
detailed physical interpretations of the infrared SED are challenging, due to 
the strong dependence on the relatively unknown radiative transfer properties 
of the grains and detailed disk structure (e.g., the inner disk radius or 
vertical scale height).  

In the early evolution stages (Class I and II), material from the inner disk 
is dragged in magnetospheric funnel flows to the stellar surface, with an 
accretion shock resulting upon impact \citep[see the review by][]{najita00}.  
This process is responsible for the observed continuum excesses 
\citep{calvet98,johns01,muzerolle03} and the shapes and strengths of emission 
lines in YSOs \citep{hartmann94,muzerolle98,muzerolle01}.  The most common 
observational measurement providing a breakdown of objects as accreting or 
non-accreting is the equivalent width ($W$) of the H$\alpha$ emission line.  
Although a standard division at $W = 10$\,\AA\ was set by historical instrument 
limitations rather than a physical motivation, this criterion provides an 
effective discriminant as many properties of weak-line (WTTSs; $W \le 10$\,\AA; 
non-accreting) and classical (CTTSs; $W > 10$\,\AA; accreting) T Tauri stars 
are remarkably different 
\citep[e.g.,][]{ghez93,osterloh95,chiang96,stelzer01}.  

Millimeter and submillimeter observations of circumstellar disks can provide 
unique information.  These observations probe the cool, outer parts of the 
disk, where giant planets are expected to form and contamination from the 
stellar photosphere is negligible.  The low submillimeter opacities in disks 
can be used to extrapolate the surface density of the outer disk into the 
inner, optically thick regions and therefore determine the total disk mass 
\citep{bscg90}.  Assuming the submillimeter emission arises in an optically 
thin, isothermal portion of the disk, the flux density ($F_{\nu}$) and disk 
mass ($M_d$) are directly proportional \citep{hildebrand83}:
\begin{equation}
M_d = \frac{d^2 F_{\nu}}{\kappa_{\nu} B_{\nu}(T_c)},
\end{equation}
where $d$ is the distance, $\kappa_{\nu}$ is the opacity, and $B_{\nu}(T_c)$ is 
the Planck function at a characteristic temperature $T_c$.  Moreover, 
observations and theoretical models of the opacity in the submillimeter 
indicate that $\kappa_{\nu}$ is well-matched by a simple power-law in frequency 
with index $\beta$, although the proposed normalizations vary significantly 
\citep{hildebrand83,wright87,pollack94,henning96}.  With the same optically 
thin, isothermal disk assumptions, the submillimeter continuum emission should 
behave roughly as $F_{\nu} \propto \nu^{2+\beta}$.  So, with major caveats (see 
\S 3 and the Appendix), a single submillimeter flux density can give the mass 
of a disk and $\ge 2$ flux points can reveal the frequency dependence of the 
opacity.  Assuming a uniform grain composition and shape, the frequency 
behavior of the opacity is set by the size distribution of the grains in the 
disk.  A number of single-dish surveys with single-element (or small arrays of) 
bolometers have been conducted in the Taurus-Auriga star-forming region to 
address these issues, most of which were carried out at 1.3\,mm 
\citep{weintraub89,bscg90,adams90,beckwith91,mannings94b,osterloh95,motte01}.
Current instrumentation provides the opportunity for significantly more 
sensitive observations of disks in the submillimeter.

High resolution observations with (sub-)millimeter interferometers have 
confirmed that circumstellar dust disks are geometrically thin with radii on 
the order of 100\,AU \citep[e.g.,][]{dutrey96,kitamura02}.  Detailed studies of 
individual disks reveal molecular gas in Keplerian rotation around the central 
star \citep[e.g.,][]{weintraub89b,koerner93a,koerner93b,dutrey94,koerner95,mannings97,duvert98,guilloteau98,simon00,corder05}.  While molecular gas is the 
primary reservoir of mass in a disk, it is difficult to directly determine 
$M_d$ from the high resolution spectral line data because the brightest, easily 
detectable lines (i.e., the rotational transitions of CO) are optically thick 
\citep{beckwith93,dutrey96} and likely to be severely depleted 
\citep{dutrey94,dutrey03}.  Interpretation of these lines and those from trace 
molecular species require sophisticated models of the disk structure 
\citep[e.g.,][]{dartois03,kamp04} and chemistry 
\citep[e.g.,][]{vanzadelhoff01,vanzadelhoff03,aikawa02,qi03}.  Despite the 
tremendous amount of information provided by these observations, our knowledge 
is still limited to a relatively few disks on account of the large amount of 
time which must be invested in an interferometric observation.

Multiwavelength submillimeter data could prove useful in placing observational 
constraints on the dominant mechanism of planet formation.  By comparing with 
infrared SEDs and diagnostics of accretion, we can investigate the dissipation 
of disks as a function of radius and see if there is consistency with the 
timescales expected from the collisional growth of planetesimals.  The 
functional form of the opacity may provide information on the mean grain size 
distribution in the disk, and therefore evidence for the growth of grains 
demanded by the standard models of planet formation 
\citep[e.g.,][]{beckwith00}.

In this paper, we present a large catalog of such data for most of the known 
YSOs in the Taurus-Auriga star-forming region.  The survey is uniform, 
sensitive, and provides the most multiwavelength measurements of the 
submillimeter continuum spectra of YSOs to date.  In \S 2 we discuss the 
observations and data reduction procedures.  In \S 3 we present a simple disk 
model and use it to derive circumstellar disk masses, place some new 
observational constraints on the submillimeter opacity properties of disks, and 
examine relationships between the disk properties and those of the central 
stars.  The results are discussed in \S 4, and our conclusions are summarized 
in \S 5.  A brief Appendix is included with a more in-depth discussion of the 
disk models we employ and comments on some particularly interesting sources.

\section{Observations and Data Reduction}

Simultaneous 450 and 850\,$\mu$m continuum photometry observations of 90 YSOs 
in the Taurus-Auriga star-forming region were obtained with the Submillimeter 
Common User Bolometer Array \citep[SCUBA:][]{holland99} at the 15\,m James 
Clerk Maxwell Telescope (JCMT) between 2004 February and 2005 January.  
Accurate reference coordinates (to $\sim$1\arcsec) for each object were 
obtained from the 2MASS Point Source Catalog.  The effective FWHM beam 
diameters for SCUBA photometry are 9\arcsec\ and 15\arcsec\ at 450 
($\lambda_{\rm{eff}} = 443$\,$\mu$m) and 850\,$\mu$m ($\lambda_{\rm{eff}} = 
863$\,$\mu$m), respectively.  The precipitable water vapor (PWV) levels in 
these observations were 1.6\,mm in the mean, corresponding to zenith opacities 
of 0.32 at 850\,$\mu$m and 1.73 at 450\,$\mu$m.  More than 50\% of the 
observations were conducted in very dry conditions (PWV $\le 1.5$\,mm).  The 
data were acquired in sets of 18\,s integrations in a small nine-point jiggle 
pattern with the secondary mirror chopping (typically) 60\arcsec\ in azimuth at 
7.8\,Hz.  Each set consisted of between 15 and 40 integrations, and each source 
was usually observed for two sets.  Frequent skydip observations were used to 
determine atmospheric extinction as a function of elevation and time.  Pointing 
updates on nearby bright standard sources were conducted between sets of 
integrations: the rms pointing offsets were $\le 2\arcsec$.  Mars and Uranus 
were used as primary flux calibrators, observed at least once per night when 
available.  The secondary calibrators HL Tau, CRL 618, and CRL 2688 were also 
observed approximately once every 60 to 90 minutes.

The demodulated SCUBA data were flatfielded, despiked, and corrected for 
extinction and residual sky emission using standard tasks in the SURF software 
package \citep{jenness98a,jenness98b}.  The ``unused" bolometers in the SCUBA 
arrays provide a distinct advantage in sky subtraction over the standard simple 
demodulation utilized for single-element (or small array) detectors.  With 
SCUBA, this technique has resulted in a factor of $\sim$3 increase in the 
signal-to-noise ratio \citep{holland99} and should give more robust flux 
measurements.  The mean and standard deviation voltages were used to determine 
the flux density and rms noise level for each source, after appropriate scaling 
based on the gain values derived from observations of the calibrators.  
Repeated observations of the flux calibrators in a given night of observing 
indicate a systematic uncertainty in these gain factors of $\sim$10\% at 
850\,$\mu$m and $\sim$25\% at 450\,$\mu$m.  These systematic errors dominate 
the uncertainties for brighter sources.  Observations of an additional 44 
sources were obtained from the SCUBA online archive and reduced in the same 
manner, accounting for the differences in filter sets for data taken before 
1999 November.  In a few cases, observations of the same source from several 
different nights were combined after the processing to yield very sensitive 
data.  For those cases, the combined data are consistent with the individual 
datasets when the increased integration time is considered.  

The Submillimeter High Angular Resolution Camera \citep[SHARC-II:][]{dowell03} 
on the 10\,m Caltech Submillimeter Observatory (CSO) telescope was also used to 
image 39 YSOs in the 350\,$\mu$m continuum between 2004 March and 2005 
January.  The FWHM beam diameter of the SHARC-II point-spread function at 
350\,$\mu$m is roughly 9\arcsec, achieved by employing an active dish surface 
optimization system at the CSO.  Due to the low atmospheric transmission at 
this wavelength, these observations were only conducted when the PWV level was 
$\le 1.6$\,mm, corresponding to 350\,$\mu$m zenith opacities of less than 
$\sim$2.  Opacity measurements at 225\,GHz were taken every 10 minutes with a 
dedicated tilting water vapor monitor observing at fixed azimuth.  The 
observations were conducted by constantly sweeping the telescope in the 
vicinity of the source in an alt-az Lissajous pattern, providing small 
Nyquist-sampled maps.  At least three separate maps were taken for each source, 
with between 120 and 600\,s of integration per map.  The aforementioned SCUBA 
calibrators were also observed every 60 to 90 minutes for pointing updates and 
flux calibration.  The SHARC-II data reduction was conducted using the CRUSH 
software package \citep{kovacs}.  Flux densities were 
measured in a circular aperture with a radius of 30\arcsec, and rms noise 
levels were determined from the background pixels.  Repeated measurements of 
standard calibration sources show that the absolute flux calibration is 
accurate to within 25\%.  

\begin{figure}
\epsscale{1.0}
\plotone{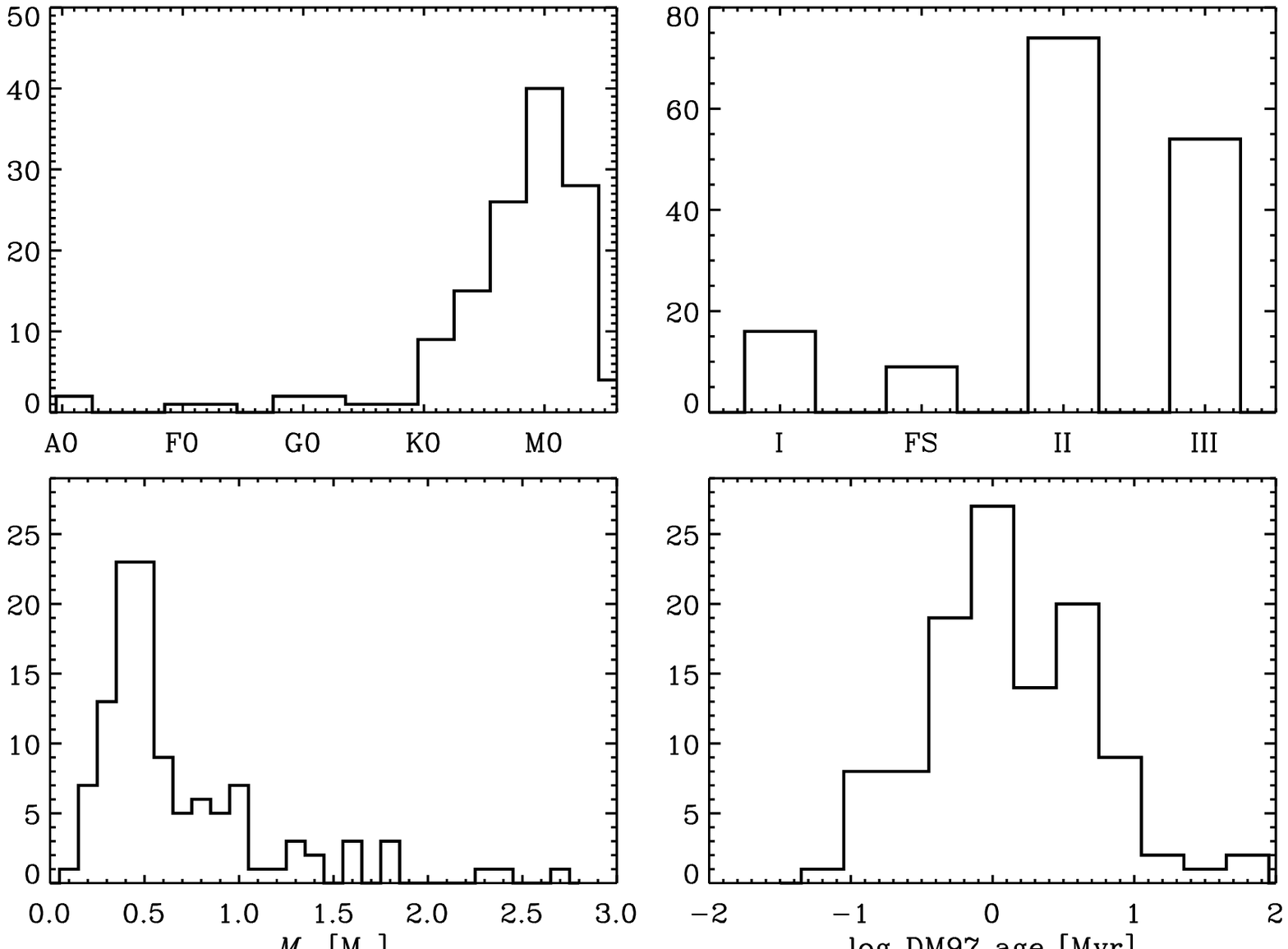}
\figcaption{Histograms highlighting some key properties of the submillimeter 
sample.  The upper left panel shows the distribution of spectral types from the 
literature (see Table \ref{results_table}), which is primarily constrained to K 
and M types.  The upper right panel marks the relative numbers of YSOs of 
various SED classifications (Class I, Flat-Spectrum, Class II, and Class III; 
see also Table \ref{results_table}): these classifications are discussed in 
detail in \S 4.  The bottom panels show the number distributions of stellar 
masses and ages, inferred as described in \S 3.4 using the \citet{dantona97} 
theoretical pre-main$-$sequence models. \label{sample}}
\end{figure}

Our sample was selected primarily from the compilation of \citet{kh95}, and was 
designed to contain roughly equal numbers of Class II and III objects, WTTSs 
and CTTSs, and single and multiple stars.  The histograms in Figure 
\ref{sample} summarize some of the key properties of the sample.  Table 
\ref{results_table} gives a collection of submillimeter properties for 153 YSOs 
in Taurus-Auriga: 90 sources with new SCUBA and SHARC-II data, 44 with archival 
SCUBA observations and SHARC-II data, 4 with data from the literature, and 15 
others with SHARC-II data and flux densities from the literature (see the table 
notes).  This table lists the 350, 450, 850\,$\mu$m, and 1.3\,mm flux densities 
(the latter from the literature) and statistical errors (1-$\sigma$ rms noise 
levels) or 3-$\sigma$ upper limits in units of mJy per beam, disk masses (see 
\S 3.2) and submillimeter continuum slopes (see \S 3.3), and various other 
relevant properties.  The projected FWHM beam diameters at the assumed distance 
of Taurus-Auriga \citep[$d = 140$\,pc;][]{elias78} are 1260\,AU for both 350 
and 450\,$\mu$m and 2100\,AU for 850\,$\mu$m.  In this paper, we assume that 
all of the sources are unresolved, and therefore the values in Table 
\ref{results_table} are actually the integrated continuum flux densities (in 
units of mJy).  This assumption is valid for Class II and III sources, where 
the submillimeter emission originates in a disk with a radius of a few hundred 
AU at most \citep[see the interferometric observations 
of][]{dutrey96,kitamura02}.  

On the other hand, submillimeter continuum maps of the Class I sources in this 
sample usually show a significant amount of extended emission from the outer 
envelope in addition to a bright, central concentration of emission (itself 
perhaps marginally resolved) from the disk and inner envelope 
\citep[e.g.,][]{chandler00,hogerheijde00,shirley00,motte01,chini01,young03}.  
The non-mapping photometry observations at 450 and 850\,$\mu$m presented here 
exclude the extended emission component, and therefore only sample the bright 
peak of emission which presumably originates from warm dust in the inner 
envelope and/or a disk.  Because of the unknown density structure of the inner 
envelope, it is not possible to unambiguously determine what fraction of this 
emission peak is contributed by a compact object (i.e., disk) without 
interferometric observations \citep[see the discussion by][and references 
therein]{young03}.  For some of these objects, there is also the possibility 
that the 60\arcsec\ chop throw would place the ``off" position in the extended 
envelope emission, and therefore the flux densities listed in Table 
\ref{results_table} could be slightly underestimated.  The reader should keep 
in mind that the submillimeter properties of Class I YSOs in this paper most 
likely refer to a combination of disk and envelope contributions.  The notes in 
Table \ref{results_table} provide references to submillimeter maps of the Class 
I YSOs in the literature when available.

\begin{figure}
\epsscale{0.55}
\plotone{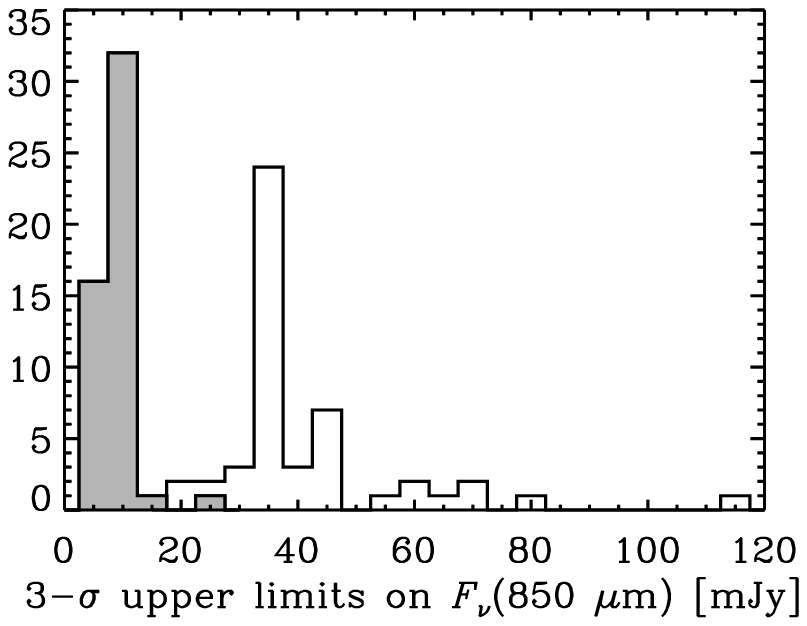}
\figcaption{The distributions of 3-$\sigma$ upper limits for the sources which 
were not detected in both this survey (\emph{filled histogram}) and the 
combined surveys of \citet{bscg90} and \citet{osterloh95} at 1.3\,mm 
(\emph{unfilled histogram}).  The upper limits are in units of mJy, and the 
1.3\,mm measurements have been scaled according to $F_{\nu} \propto \nu^2$ (see 
\S 3.3) to enable direct comparison with the 850\,$\mu$m measurements.  The 
submillimeter survey presented here is roughly a factor of 5 more sensitive in 
terms of flux density limits, and is considerably more uniform. 
\label{noise_comparison}}
\end{figure}

The primary observational goal of this survey was to take advantage of the 
stability and efficiency of the SCUBA instrument to obtain a 850\,$\mu$m sample 
with a relatively uniform flux density limit of $\sim$10\,mJy (3-$\sigma$).  
The mean 3-$\sigma$ upper limit for undetected sources at 850\,$\mu$m in this 
survey is 8.4\,mJy (the median is the same), with a standard deviation in the 
upper limits of 3.1\,mJy.  For comparison, the same sources in the combined 
1.3\,mm surveys in Taurus-Auriga conducted by \citet{bscg90} and 
\citet{osterloh95} have a mean 3-$\sigma$ upper limit of 19\,mJy (median of 
16\,mJy) and a standard deviation in the upper limits of 10\,mJy.  If we assume 
that the submillimeter continuum emission behaves as $F_{\nu} \propto \nu^2$ 
(see \S 3.3), then a factor of 2.3 can be used to scale the 1.3\,mm 
measurements with those at 850\,$\mu$m.  The resulting scaled 1.3\,mm mean 
upper limit is then 44\,mJy (median of 37\,mJy).  The distributions of the 
upper limits of undetected sources are shown in Figure \ref{noise_comparison}.  
In terms of flux density limits on undetected sources, our survey is roughly a 
factor of 5 more sensitive than previous single-dish work and is also 
considerably more uniform.  The distributions of the signal-to-noise ratios for 
detected sources in the various surveys are similar, although there are 
generally higher ratios at 850\,$\mu$m.  For the sources common to the 
850\,$\mu$m and 1.3\,mm samples, the detection rates are $64 \pm 7$\% and $47 
\pm 6$\%, respectively.  

\section{Results}

\subsection{A Simple Disk Model}

A model of the submillimeter continuum emission is needed to extract physical 
information (e.g., disk masses) from the data.  In order to incorporate some 
non-negligible optical depth and a radial temperature distribution, the 
simplistic methods outlined in \S 2 (e.g., Equation 1) are passed over in favor 
of one that fits the disk SED with a power-law structural model 
\citep[see][]{adams87,bscg90}.  In this scheme, the SED (from the mid-infrared 
through the submillimeter) is assumed to be generated from thermal reprocessing 
of starlight by a geometrically thin dust disk, with the flux density given by
\begin{equation}
F_{\nu} = \frac{\cos{i}}{d^2} \int_{r_{\circ}}^{R_d} B_{\nu}(T_r) \left( 1 - e^{-\tau_{\nu,r}\sec{i}} \right) 2 \pi r dr
\end{equation}
where $i$ is the inclination angle, $r_{\circ}$ the inner radius, $R_d$ the 
outer radius, $B_{\nu}(T_r)$ the Planck function at a radius-dependent 
temperature, and $\tau_{\nu,r}$ the optical depth of the disk 
material.\footnote{Because Equation 2 implicitly assumes a constant source 
function in the disk, it is only a valid approximation when the inclination 
angle is not too large.  A more sophisticated treatment of radiative transfer 
is required for nearly edge-on disks \citep[e.g.,][]{chiang99}.}  In essence, 
the flux density is computed by summing the thermal emission from a continuous 
set of dust annuli weighted by the radiative transfer properties of the 
material.  The radial temperature distribution is taken to be a power law 
\begin{equation}
T_r = T_1 \left( \frac{r}{1\:\rm{AU}} \right)^{-q}
\end{equation}
where $T_1$ is the temperature at $r = 1$\,AU.  The optical depth is the
product of the disk opacity, $\kappa_{\nu}$, and the radial surface density
profile, $\Sigma_r$, which is also taken to be a power law:
\begin{equation}
\Sigma_r = \Sigma_{\circ} \left( \frac{r}{r_{\circ}} \right)^{-p}.
\end{equation}
We assume that the opacity is a power law in frequency with index $\beta$ and a 
normalization of 0.1\,cm$^2$ g$^{-1}$ at 1000\,GHz \citep{bscg90}.  This value 
assumes a 100:1 mass ratio between gas and dust.

Because a given disk typically has relatively few SED datapoints, fitting the 
SED with the model described above requires that some of the remaining 8 
parameters ($i$, $r_{\circ}$, $R_d$, $\Sigma_{\circ}$, $p$, $T_1$, $q$, 
$\beta$) be fixed.  Fortunately, the precise values of $i$, $r_{\circ}$, and 
$R_d$ do not significantly affect the determination of interesting physical 
parameters as long as they lie in a realistic range.  A fiducial set of fixed 
parameters is adopted here: $i = 0\degr$, $r_{\circ} = 0.01$\,AU, $R_d = 
100$\,AU, and $p = 1.5$.\footnote{See the Appendix for a more detailed 
examination of the effects of various parameter choices.}  The inner and outer 
disk radii are typical values based roughly on the dust sublimation temperature 
\citep[e.g.,][]{dullemond01,muzerolle03} and direct disk size measurements 
\citep[e.g.,][]{dutrey96,kitamura02,akeson05}.  The surface density index, $p$, 
is the most difficult parameter to constrain observationally.  The value 
selected here is obtained when the compositions of the planets in the solar 
system are augmented to cosmic abundances and smeared out into annuli: the 
Minimum Mass Solar Nebula \citep[MMSN;][]{weidenschilling77}.  The inclination 
value is set merely as a computational convenience.  The remaining parameters 
($\Sigma_{\circ}$, $T_1$, $q$, $\beta$) must be determined from the data.  

\subsection{Disk Masses}

Submillimeter continuum observations provide measurements of disk masses.  
However, the simplistic conversion of a flux density into a mass via Equation 1 
masks some important complications.  For example, $M_d$ could be uncertain to a 
factor of $\sim$2 due to its roughly linear relationship with $T_c$.  More 
fundamentally, the relationship between $F_{\nu}$ and $M_d$ is nonlinear due to 
the significant fraction of the submillimeter emission which is optically 
thick \citep[e.g.,][]{bscg90}.  By assuming optically thin emission and using 
Equation 1, $M_d$ could be underestimated (particularly for objects with larger 
flux densities).  To avoid these problems and fit the SEDs with the model 
described by Equation 2, mid- and far-infrared flux densities were taken from 
the \emph{IRAS} Point Source Catalog and the compilations of \citet{weaver92} 
and \citet{kh95}.  Using data at shorter wavelengths ($\lambda \lesssim 
5$\,$\mu$m) runs the risk of contamination from an extincted photosphere, and 
therefore introduces more parameters into the problem (e.g., effective 
temperature, extinction, stellar radius).  The submillimeter data presented 
here were supplemented whenever possible with flux densities from the 
literature \citep{adams90,bscg90,beckwith91,mannings94b,osterloh95,motte01}.  
We adopted absolute flux calibration uncertainties of 20\% in the infrared and 
25\% ($\lambda \le 800$\,$\mu$m) or 20\% ($\lambda > 800$\,$\mu$m) in the 
submillimeter.  Systematic and statistical errors were combined for each 
individual flux density measurement.  

\begin{figure}
\epsscale{0.5}
\plotone{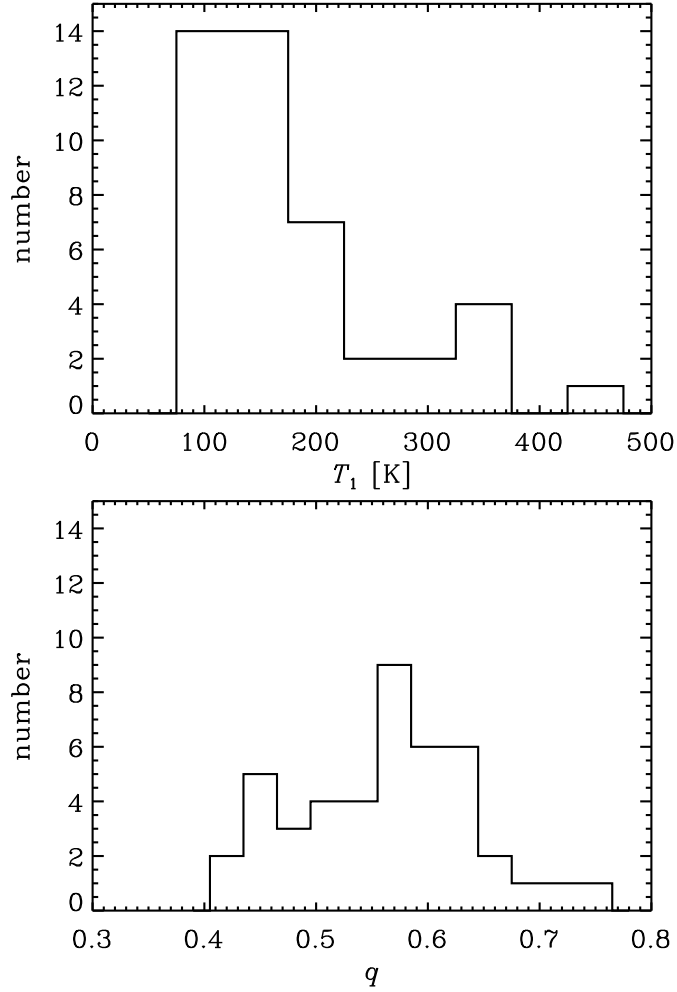}
\figcaption{The distributions of the best-fit disk model parameters of the 
normalization ($T_1$: \emph{top}) and power law index ($q$: \emph{bottom}) of 
the radial temperature profile.  The mean and median values for the sample are 
$0.56 \pm 0.08$ and $0.58$ for $q$, and $178 \pm 85$\,K and $148$\,K for 
$T_1$ (quoted errors are standard deviations).  \label{qTdists}}
\end{figure}

The disk mass and opacity index, $\beta$, are strongly coupled parameters, 
making it difficult to independently infer their values 
\citep[see][]{beckwith91}.  Observations and models of interstellar grains in 
the molecular ISM, where the material is still diffuse enough to safely assume 
optically thin thermal emission, indicate that $\beta \approx 2$ 
\citep{erickson81,schwartz82,draine84}.  However, different mineralogies or 
grain size distributions in a disk could decrease the index down to $\beta \sim 
0$ \citep[e.g.,][]{pollack94}.  Because of the uncertainties in independently 
measuring $\beta$ and $M_d$, we modeled individual SEDs for various values of 
$\beta$ (between 0 and 2), as well as the typical compromise value for disks, 
$\beta = 1$.  Values of $T_1$, $q$, and $\Sigma_{\circ}$ (note that for this 
model $M_d/\rm{M}_{\odot} \approx 2 \times 10^{-35} \Sigma_{\circ}/\rm{g \,\,
cm}^{-2}$) for 44 objects in the sample were determined by fitting the SEDs to 
Equation 2 and minimizing the $\chi^2$ statistic.  Table \ref{fit_results} 
gives the results of these fits for $\beta = 1$, including the reduced $\chi^2$ 
values ($\tilde{\chi}_{\nu}^2$), degrees of freedom in the fit ($\nu$), and 
references for the infrared and submillimeter SED data from the literature.  
Much more sophisticated disk models \citep{menshchikov99,chiang01,semenov05} 
predict disk masses for a few of the same sources which are within a factor of 
2-3 of those presented here.  Figure \ref{qTdists} shows the distributions of 
the best-fit values of $q$ and $T_1$.  The mean values of $q$ and $T_1$ are 
$0.56 \pm 0.08$ and $178 \pm 85$\,K, respectively (quoted errors are standard 
deviations of the distributions).  We define the ``median disk model" to have 
the above set of fiducial parameters and the median values $q = 0.58$ and $T_1 
= 148$\,K, as well as $\beta = 1$.  Due to the high optical depths at infrared 
wavelengths compared to the submillimeter, the parameters of the temperature 
profile, $T_1$ and $q$, are often not strongly affected by changes in $\beta$. 

\begin{figure}
\epsscale{0.55}
\plotone{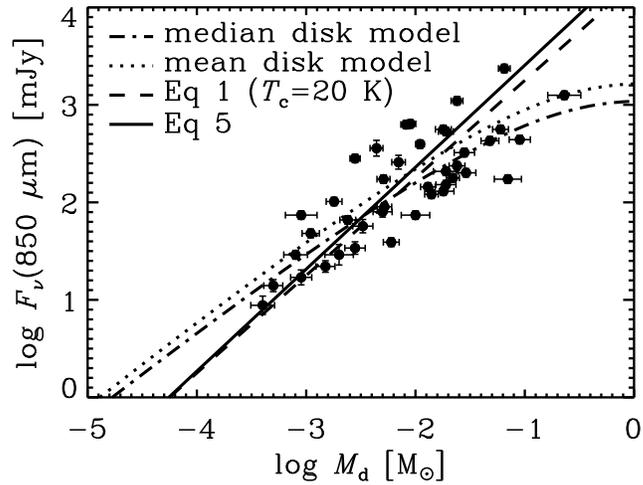}
\figcaption{The relationship between the disk mass and 850\,$\mu$m flux density 
for the sources in Table \ref{fit_results}.   The solid line is the best linear 
fit to the data in log-log space, given as Equation 5 in the text.  The dashed 
line is the best fit to the data based on the optically thin, isothermal disk 
model given in Equation 1: the best-fit characteristic temperature is $T_c = 
20$\,K.  The dotted and dash-dotted curves show the mean and median disk model 
behaviors, as described in the text.  Although a considerable dispersion exists 
(the rms residual dispersion around the best-fit power law model given in 
Equation 5 is 0.2\,dex), the apparent correlation here permits an empirical 
calibration of the $M_d$ $-$ $F_{\nu}$ relationship.  \label{Md_F850}}
\end{figure}

The SEDs of most of the YSOs in the survey sample were not fitted as described 
above because they either lack data (i.e., there were too few degrees of 
freedom), are undetected in the submillimeter, or have SEDs which indicate such 
a simple model is insufficient.  However, the results of the SED fitting can be 
used to determine an empirical conversion between a submillimeter flux density 
and a disk mass.  In Figure \ref{Md_F850} we show the relationship between the 
850\,$\mu$m flux densities and the best-fit values of $M_d$ (for $\beta = 1$) 
from the SED fitting.  This relationship is well described by a simple power 
law,
\begin{equation}
\frac{M_d}{\rm{M}_{\odot}} = (5 \pm 2) \times 10^{-5} \left[ \frac{F_{\nu}(850\,\mu\rm{m})}{\rm{mJy}} \right]^{0.96 \pm 0.03} 
\end{equation} 
which is shown as a solid line in Figure \ref{Md_F850}.  A fit of the same data 
to Equation 1 (assuming the same opacity function given above, where 
$\kappa_{\nu} = 0.035$\,cm$^2$ g$^{-1}$ at 850\,$\mu$m) is shown as a dashed 
line, and gives a best-fit characteristic temperature $T_c = 20$\,K.  For the 
median disk model, this value of $T_c$ occurs at a disk radius of approximately 
30\,AU.  Also shown are the relationships between $F_{\nu}$ and $M_d$ for the 
mean and median disk models.  Disk mass values and upper limits for the objects 
which were not fitted with these models were computed from Equation 5.  For 
sources without 850\,$\mu$m measurements which could not be fitted with a disk 
model, a similar analysis as above was used to derive values of $M_d$ from the 
1.3\,mm flux density: $M_d/{\rm M}_{\odot} \approx 10^{-6} 
(F_{\nu}/\rm{mJy})^{1.5}$.  Disk masses (or 3-$\sigma$ upper limits) are 
included in Table \ref{results_table} for all of the sources in the survey 
sample.  Those masses, which use $\beta = 1$, will be adopted throughout this 
paper, unless specifically mentioned otherwise.  Our fitting results show that 
the systematic errors in the disk mass due to the \emph{a priori} unknown value 
of $\beta$ are $\pm$0.5\,dex on average, or a factor of 3, for a reasonable 
range of $\beta$ (0 to 2).  The $M_d$ values inferred for Class I objects 
should be considered only as upper limits on the disk mass, as there is likely 
a flux contribution from the inner envelope.

\begin{figure}
\epsscale{0.5}
\plotone{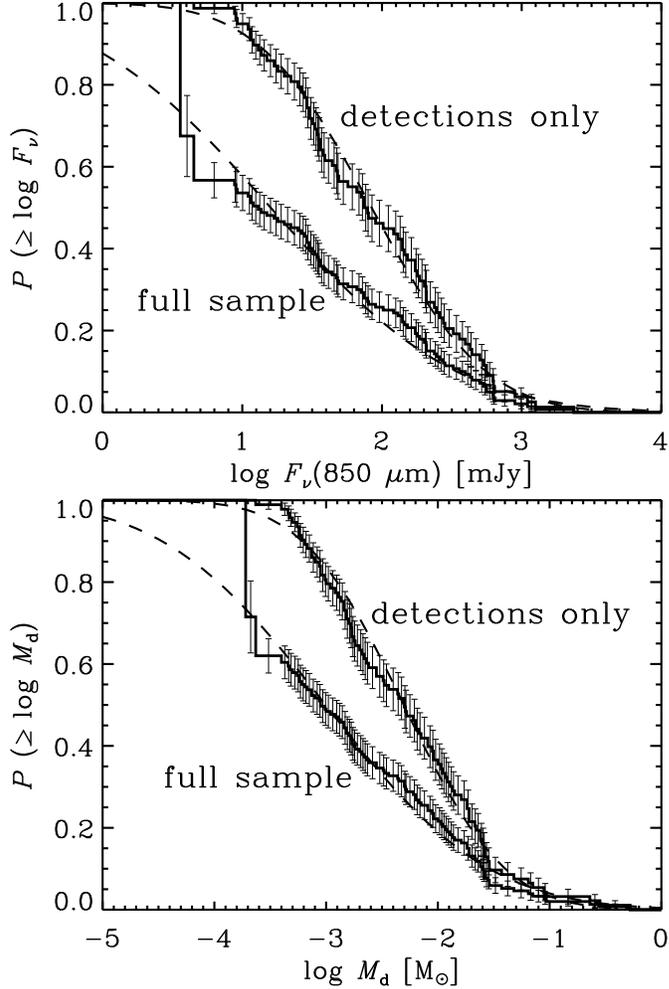}
\figcaption{Cumulative distributions of 850\,$\mu$m flux densities for 140 
objects in the full sample and the 78 of those which have $\ge 3$-$\sigma$ 
detections (\emph{top}) and disk masses for 153 objects in the full sample and 
93 objects which have submillimeter detections (\emph{bottom}).  The full 
sample distribution functions were computed using the Kaplan-Meier estimator to 
incorporate 3-$\sigma$ upper limits.  The ordinate values represent the 
probability that an object in the sample has a flux density or disk mass 
greater than or equal to each abscissa value.  The flux densities are 
log-normally distributed: the full sample with mean $1.20 \pm 0.02$ (16\,mJy) 
and variance $1.08 \pm 0.06$\,dex and the detections subsample with mean $1.93 
\pm 0.01$ (85\,mJy) and variance $0.41 \pm 0.02$\,dex.  The same distribution 
holds for the disk masses: the full sample with mean $-3.00 \pm 0.02$ 
($10^{-3}$\,M$_{\odot}$) and variance $1.31 \pm 0.06$\,dex and the detections 
subsample with mean $-2.31 \pm 0.01$ ($5 \times 10^{-3}$\,M$_{\odot}$) and 
variance $0.50 \pm 0.02$\,dex.  The best-fit distributions are overlaid as 
dashed curves.  \label{F850_CDF}}
\end{figure}

Figure \ref{F850_CDF} shows the cumulative distributions of the 850\,$\mu$m 
flux densities and disk masses.  This figure shows the distributions of the 
full sample and a subsample consisting of only those sources which have a $\ge 
3$-$\sigma$ detection at a submillimeter wavelength.  The ordinates in these 
plots are defined as the probability of an object having a value equal to or 
greater than the abscissae.  In both figures, the Kaplan-Meier product limit 
estimator is used to construct the cumulative distributions for the full 
sample.\footnote{Application of the Kaplan-Meier estimator to the flux density 
data may be inappropriate.  Because each object was observed either until it 
was detected or a rather uniform flux density limit was reached, the function 
which describes the censoring of these data is not random.  Nevertheless, any 
effects of using the Kaplan-Meier estimator should only be noticed for flux 
densities below the completeness limit ($\sim$10\,mJy).  Because 
multiwavelength SED data were used in determining $M_d$ (and various SED 
morphologies can result in identical values of $M_d$), the censoring function 
in that case should be randomized, and therefore the use of the Kaplan-Meier 
estimator is valid.}  This method allows the incorporation of the 3-$\sigma$ 
upper limits of the $F_{\nu}$ and $M_d$ values in the full sample.  The 
computations of probabilities and their errors were conducted with the ASURV 
Rev. 1.2 software package \citep{lavalley90}, following the formalism 
introduced by \citet{feigelson85}.  A significant caveat with these cumulative 
distributions is that there is no means to account for the uncertainties in the 
values of $F_{\nu}$ and $M_d$.

Based on the detections subsample distribution in Figure \ref{F850_CDF} (not 
incorporating upper limits), we estimate the completeness limit of the survey 
to be roughly 10\,mJy.  A log-normal distribution of $F_{\nu}$ with mean $1.20 
\pm 0.02$ (16\,mJy) and variance $1.08 \pm 0.06$\,dex provides a good fit to 
the data for the full sample, whereas a mean of $1.93 \pm 0.01$ (85\,mJy) and a 
variance of $0.41 \pm 0.02$\,dex are appropriate for the detections subsample.  
The distribution of the detections subsample in Figure \ref{F850_CDF} indicates 
that 37\% of the YSOs have $M_d \ge 0.01$\,M$_{\odot}$, roughly the total mass 
of the MMSN \citep{weidenschilling77}.  Approximately 79\% of the same 
subsample have disks with masses greater than that of Jupiter.  As would be 
expected from the relationship between $F_{\nu}$ and $M_d$ discussed above, the 
disk masses are also log-normally distributed: the full sample with mean $-3.00 
\pm 0.02$ ($10^{-3}$\,M$_{\odot}$) and variance $1.31 \pm 0.06$\,dex, 
and the detections subsample with mean $-2.31 \pm 0.01$ ($5 \times 
10^{-3}$\,M$_{\odot}$) and variance $0.50 \pm 0.02$\,dex. 

\subsection{Submillimeter Continuum Slopes}

The slope of the submillimeter continuum emission from a circumstellar disk is 
empirically well-described by a simple power law in frequency: $F_{\nu} \propto 
\nu^{\alpha}$.  If the emission is assumed to be optically thin and roughly 
isothermal, Equation 2 can be written $F_{\nu} \propto B_{\nu}(T) \tau_{\nu} 
\propto \nu^{2+\beta}$ (in the Rayleigh-Jeans limit).  However, the 
submillimeter continuum has a significant contribution from optically thick 
emission originating in the dense, inner disk which causes a substantial 
deviation from the $\alpha = 2 + \beta$ relationship inferred for the optically 
thin case \citep{bscg90,beckwith91}.  We have combined the data presented here 
with additional flux densities from the literature 
\citep{adams90,bscg90,beckwith91,mannings94b,osterloh95,motte01} to determine 
the values of $\alpha$ given in Table \ref{results_table}.  For objects with 
more than 2 submillimeter flux densities, $\alpha$ was measured from a linear 
fit in the $\log{\nu}$-$\log{F_{\nu}}$ plane.  When only 1 or 2 flux densities 
were available, values or 3-$\sigma$ upper limits of $\alpha$ were determined 
from a simple spectral index.  

\begin{figure}
\epsscale{1.0}
\plotone{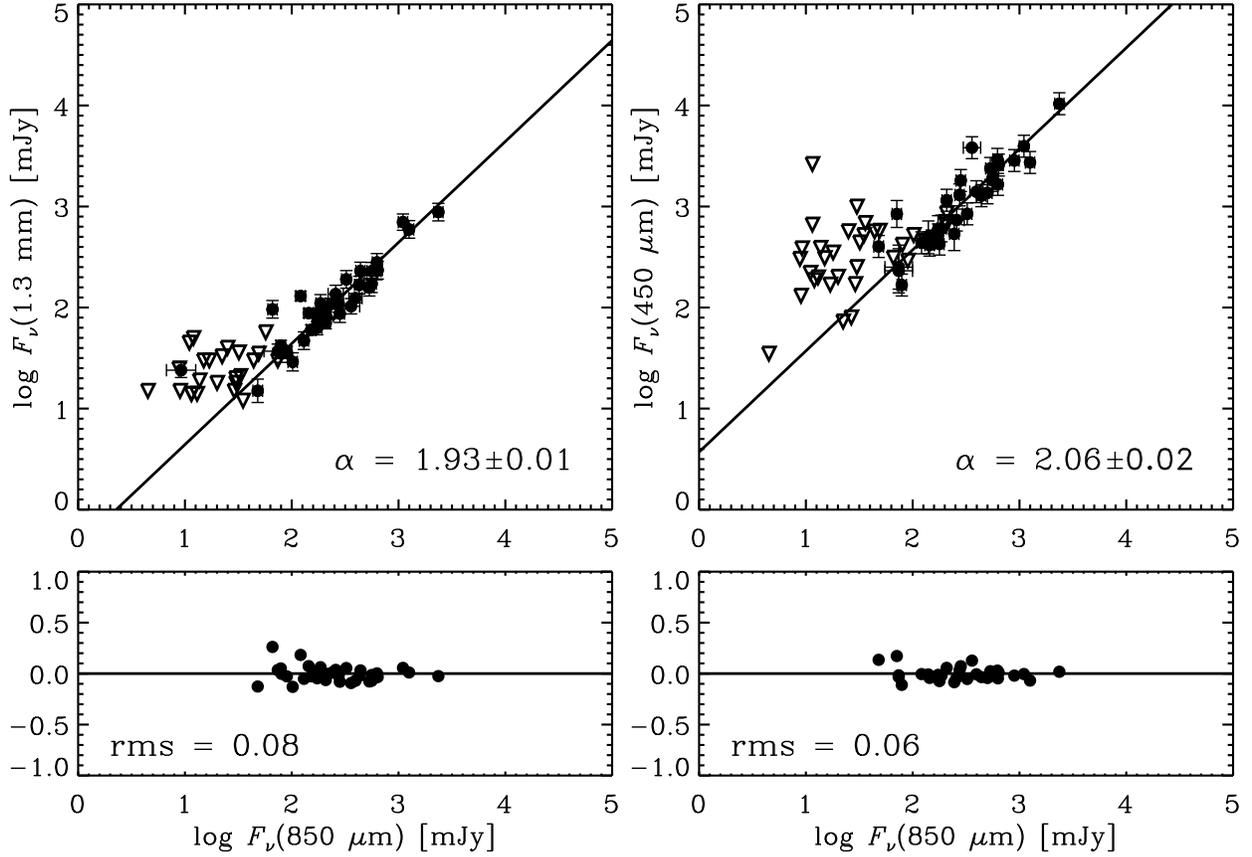}
\figcaption{The upper panels show 1.3\,mm (\emph{left}) and 450\,$\mu$m 
(\emph{right}) flux densities plotted against 850\,$\mu$m flux densities.  The 
open triangles mark 3-$\sigma$ upper limits.  For sources with no obvious error 
bars, errors are smaller than the symbol size.  The solid lines are power laws 
of the form $F_{\nu} \propto \nu^{\alpha}$, and the best-fit value of $\alpha$ 
is given in the lower right corner (determined from the intercept of a linear 
fit in log-log space).  The fits are only conducted for 
$\log{F_{\nu}(850\,\mu\rm{m})} > 1.5$, and sources with anomalous continuum 
slopes are excluded (see Table \ref{results_table}).  The bottom panels give 
the residuals to the fits and the rms values of the residuals.  Considering the 
scatter around the best-fit relationships, the data indicate that the continuum 
slope is roughly the same regardless of the wavelength range in which it is 
measured.  \label{fratios}}
\end{figure}

Figure \ref{fratios} compares the submillimeter continuum slopes from 450 to 
850\,$\mu$m and 850\,$\mu$m to 1.3\,mm.  For the full sample, the best-fit 
slopes are $\alpha = 2.06 \pm 0.02$ and $1.93 \pm 0.01$ for the two wavelength 
regions, respectively.  Because the submillimeter continuum emission is 
generated in the outer disk where temperatures are low, the Rayleigh-Jeans 
limit criterion is not satisfied (because $h\nu \sim kT$) and the continuum 
slope at shorter wavelengths (nearer to the peak of the thermal emission) could 
be systematically smaller than at longer wavelengths.  However, such an effect 
is not seen in Figure \ref{fratios}: in fact, the shorter wavelength slope is 
slightly steeper than at longer wavelengths.  This implies that the shape of 
the submillimeter continuum is not set by the grain temperature distribution 
alone, but also by the amount of optically thick emission and/or the spectral 
behavior of the opacity function.  Within the uncertainties, it does not 
significantly matter where in the submillimeter continuum the slope is measured 
(at least between 350\,$\mu$m and 1.3\,mm).  The slightly shallower best-fit 
slope for the longer wavelength data may simply be noise, or could be caused by 
a small fraction of the 1.3\,mm data which are contaminated by non-disk 
emission: e.g., free-free or gyrosynchrotron radiation from a wind or outflow 
\citep[e.g.,][]{chiang96}.  Another possibility is a real concavity to the 
long-wavelength SEDs: the models of \citet{pollack94} predict a steeper opacity 
function shortward of $\sim$650\,$\mu$m.  

\begin{figure}
\epsscale{0.55}
\plotone{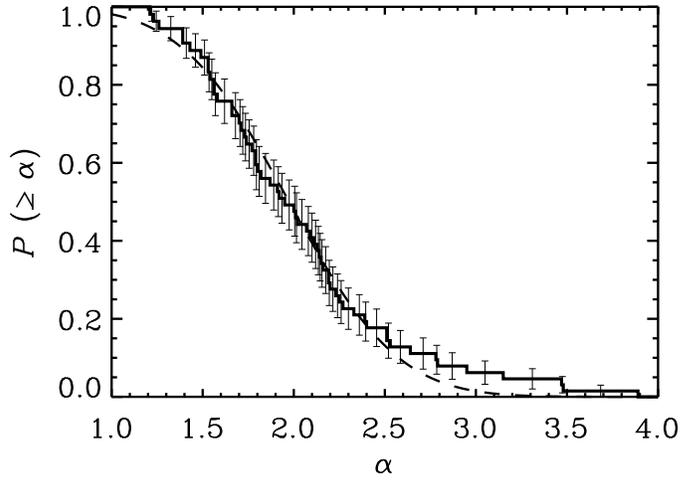}
\figcaption{The cumulative distribution of submillimeter continuum slopes 
($\alpha$: defined by $F_{\nu} \propto \nu^{\alpha}$) for 84 YSOs, computed 
using the Kaplan-Meier estimator to incorporate 3-$\sigma$ upper limits.  The 
ordinate values represent the probability that an object in the sample has a 
continuum slope greater than or equal to $\alpha$.  The continuum slopes are 
normally distributed with mean $1.97 \pm 0.01$ and variance $0.22 \pm 0.02$, 
overlaid with a dashed curve.  The high slope tail of the distribution is 
slightly enhanced.  \label{alpha_CDF}}
\end{figure}

The cumulative distribution of $\alpha$ is shown in Figure \ref{alpha_CDF}, 
constructed using the Kaplan-Meier estimator to incorporate 3-$\sigma$ upper 
limits.  The median value of $\alpha$ is 2.0, while only 6\% of the sample has 
$\alpha \ge 3$, a typical value adopted in the literature due to the 
(incorrect) assumption of optically thin emission in the Rayleigh-Jeans limit 
with $\beta = 1$.  A normal distribution of $\alpha$ with a mean of $1.97 \pm 
0.01$ and a variance of $0.22 \pm 0.02$ provides a decent fit to the data, but 
there is a slightly enhanced probability of larger $\alpha$.  Because some of 
the emission is optically thick, there is no straight-forward means of 
associating these values of the continuum slope with power law indices of the 
opacity function ($\beta$).  One approach is to allow $\beta$ to vary in the 
disk SED models and fit it as an additional parameter 
\citep[e.g.,][]{beckwith91,mannings94b,dutrey96}.  However, in many cases this 
severely limits the number of degrees of freedom in the fits (see Table 
\ref{fit_results}), which already make a number of assumptions.  
\citet{beckwith91} provide a means of relating $\alpha$ and $\beta$ 
analytically from other parameters in the SED model fits which essentially 
indicate that $\beta \propto \alpha$, although there is a constant offset (see 
the Appendix).  
  
Figure \ref{ab_Md} shows the measured values of $\alpha$ as a function of 
$\log{M_d}$.  The shaded region on this diagram marks the functional form of 
$\alpha(M_d)$ for $\beta = 2$ which is representative of the \emph{complete} 
range in the measured radial temperature distributions (see Figure 
\ref{qTdists}).  The disk mass values are those for $\beta = 1$, and the error 
bar shown to the lower left demonstrates the systematic uncertainty introduced 
by varying $\beta$ between 0 and 2.  All else being equal, the disk mass is 
roughly inversely proportional to the opacity, and since larger values of 
$\beta$ give lower opacities for a fixed frequency, a larger $\beta$ will also 
give a larger $M_d$ (assuming the normalization of the opacity function is 
fixed).  Even allowing for such uncertainties in the disk mass and the range of 
temperature profiles, Figure \ref{ab_Md} indicates that many of the disks in 
the sample have $\beta < 2$ (the curves showing the relationship between 
$\alpha$ and $M_d$ for lower values of $\beta$ always fall in or below the 
shaded strip: see the Appendix).  This suggests that the large optical depths 
in the disk do not completely explain the shallow measured continuum slopes, 
but that there is also an evolutionary change in the typical opacity properties 
of dust grains from the ISM (where $\beta = 2$) to a disk.  This result has 
also been noted in other studies \citep{beckwith91,mannings94b,dutrey96,dent98} 
for various different sizes and types of samples.  It is tempting to conclude 
that the apparently diminished values of $\beta$ in these disks are due to the 
collisional growth of dust grains, a necessary condition in any planet 
formation model \citep{mizuno80,pollack96}.  Models of the process indicate 
that grain growth can decrease $\beta$ to values as low as zero 
\citep[e.g.,][]{miyake93,henning95}.  The data in Figure \ref{ab_Md} also 
clearly show that there is no correlation between the disk mass and 
submillimeter continuum slope.

The coupling of $M_d$, $\beta$, and the opacity normalization makes it 
difficult to definitively associate low values of the submillimeter continuum 
slope with decreased opacity indices \citep{beckwith91}.  The actual value of 
$\kappa_{\nu}$ is the main uncertainty in the conversion of a submillimeter 
flux density into a disk mass.  Aside from the effect of the grain size 
distribution (thus the interest in grain growth), both the normalization and 
$\beta$ depend strongly on the mineralogical composition of the grains 
\citep{pollack94,henning96} and their physical shapes \citep[e.g., spherical, 
fractal, etc.; see][]{wright87}.  Further discussion of these uncertainties is 
given by \citet{bscg90} and \citet{beckwith00}.  As an example, 
\citet{wright87} indicates that $\kappa_{\nu}$ can be roughly an order of 
magnitude higher for fractal grains compared to spheres at a wavelength of 
1\,mm (see his Figure 6).  

Observational and theoretical uncertainties obfuscate the relationship between 
a measured submillimeter continuum slope and the functional form of the opacity 
in a disk.  Overcoming these difficulties to pursue evidence of the collisional 
agglomeration of dust grains in the earliest stages of planet formation will at 
least require better observations, including resolved images at wavelengths 
extending beyond $\sim$1\,mm \citep[where the emission is more optically thin; 
e.g.,][]{testi01}, flux measurements near the SED turnover point (in the 100 to 
300\,$\mu$m range), and studies of solid-state dust emission features in the 
mid-infrared \citep[e.g.,][]{vanboekel04}.  However, our results leave no doubt 
that the measured submillimeter continuum slopes for YSOs are significantly 
less than those noted for molecular clouds, where $\alpha \approx 4$.

\subsection{Connections to Stellar Properties}

The physical properties of a young star and its circumstellar disk could be 
related due to their mutual formation and subsequent gravitational and thermal 
links.  Generally, observational indications of any such relationship are 
absent, presumably due to a wide range of initial circumstellar conditions for 
individual sources and the relatively small ranges of stellar masses and ages.  
As with \citet{bscg90}, we do not find any correlations between measured 
submillimeter properties and any characteristic of the stellar photosphere 
(e.g., effective temperature, luminosity, optical fluxes or colors, etc.).  One 
perhaps notable exception is the large value of $T_1$ derived for the two A 
stars in this sample (AB Aur and V892 Tau).  \citet{natta01} also suggest that 
hotter stars have generally higher dust temperatures in their disks, but there 
is no noticeable trend for the cooler (K and M) majority of this sample.  This 
disconnect between the stellar photosphere and the outer disk, where the 
submillimeter emission is generated, is not surprising.  Despite the increase 
in vertical scale height of the disk with radius expected from hydrostatic 
equilibrium \citep[e.g.,][]{kenyon87}, radiative transfer models for 
structurally realistic disks indicate that the bulk of the submillimeter 
emission comes from the dust near the disk midplane, and not in the flared 
atmosphere which can be more directly affected by the stellar photosphere 
\citep{chiang97,chiang99}.  

The gravitational link between a young star and its disk suggests that the 
stellar and disk masses may be related.  Circumstellar disks are 
self-gravitationally stable if their mass is less than a fraction (a few 
tenths) of the stellar mass \citep{shu90,laughlin94}.  In principle, this could 
allow more massive stars to harbor more massive disks.  \citet{natta01} combine 
interferometric measurements of disks around early-type stars with the 1.3\,mm 
survey of \citet{bscg90} and claim a marginal correlation between the disk mass 
and stellar mass ($M_{\ast}$) over 2 orders of magnitude in $M_{\ast}$ for 
roughly 100 objects \citep[however, see][]{mannings00}, although the dispersion 
is substantial.  To revisit this issue, we have collected optical/near-infrared 
magnitudes and spectral classifications from various sources in the literature 
\citep[][spectral types are listed in Table \ref{results_table}]{coku79,jones79,slutskii80,herbig86,herbig88,strom89,hartmann91,gomez92,bouvier93,briceno93,hartigan94,martin94,kh95,hernandez04,white04}.  A consistent set of visual 
extinctions was determined from the $(V-I)$ color excesses, using the intrinsic 
colors tabulated by \citet{kh95} and the interstellar extinction law derived by 
\citet{cohen81}.  De-reddened visual magnitudes and spectral types were 
converted to bolometric luminosities and effective temperatures again using the 
intrinsic values of \citet{kh95}.  Stellar masses and ages were determined by 
reference to theoretical pre-main$-$sequence evolution tracks and isochrones 
\citep{dantona97} in a Hertzsprung-Russell diagram.

As Figure \ref{starscat} demonstrates, there are no correlations between the 
submillimeter properties listed in Table \ref{results_table} and stellar mass 
or age, but the ranges of those stellar properties (see Figure \ref{sample}) 
may be too limited to infer a direct evolutionary sequence.  However, the upper 
right panel of this figure shows that the region corresponding to higher mass 
disks at late times ($\ge 6$\,Myr) is significantly depopulated.  While there 
are not many objects in Taurus-Auriga with such ages, this unoccupied region in 
the diagram is consistent with other studies that indicate disk fractions 
approaching zero in the 6 to 10 Myr age range \citep[e.g.,][]{haisch01}.  
Figure \ref{CDF_MDMS} shows the cumulative distributions of the mass ratio of 
disk to star, constructed with the Kaplan-Meier estimator.  Log-normal 
distributions provide poor fits in this case, but these distributions are fit 
fairly well with power laws of index between $-1.5$ and $-2$ for mass ratios 
larger than $\sim$10$^{-3}$.  The median disk to star mass ratio is 0.5\%.  The 
fraction of disks which may be self-gravitationally unstable (mass ratios 
larger than $\sim$0.1) is negligible in Taurus-Auriga: roughly 6\%, which 
itself may be an overestimate due to envelope emission for some of the Class I 
objects at the high end of the distribution.  However, if $\beta = 2$ is more 
appropriate, then the fraction of unstable disks can be as high as one third.  
A small fraction of objects (a few percent) has a mass ratio less than 
$10^{-3}$.    

\subsection{The Effects of Multiplicity}

The evolution of circumstellar disks can be dictated by either internal (e.g., 
viscous accretion, gravitational instability, planet formation) or external 
processes.  Examples of the latter include ultraviolet photoevaporation in the 
vicinity of a massive star \citep[e.g.,][]{johnstone98}, dynamical interactions 
with other stars in a crowded cluster environment \citep{kroupa95,boffin98} or 
in a local multiple star system.  In the low stellar density Taurus-Auriga 
region, which is devoid of stars earlier than A0, the dominant external process 
affecting disk evolution is expected to be dynamical star-disk or disk-disk 
interactions in multiple star systems.  Most young stars in nearby clusters and 
main-sequence stars in the field are in multiple systems, and the multiplicity 
fraction in Taurus-Auriga may be exceptionally large 
\citep{mathieu94,mathieu00}.  The similar multiplicity fractions for YSOs and 
main-sequence field stars indicates that binary formation occurs early in 
stellar evolution (\emph{at least} before the Class II stage, and likely much 
earlier), when significant circumstellar material is still present.  
Gravitational interactions are expected to severely affect the structural 
integrity of disks in the system, including truncation of the outer parts of 
individual circum\emph{stellar} disks, gap formation in circum\emph{binary} 
disks, or even complete dissipation of circumstellar material via accretion or 
ejection \citep{artymowicz94}.  

As an example, consider a young binary system with semimajor axis $a$ and 
eccentricity $e$, which also harbors two individual circumstellar disks and a 
larger circumbinary disk.  Simulations of the gravitational dynamics in such a 
system indicate that the circumstellar disks will be truncated and a gap will 
open in the circumbinary disk, at radii which are determined primarily by the 
values of $a$ and $e$ \citep{artymowicz94}.  The disk model described in \S 3.1 
can be adjusted to determine the effects on the SED of such disk configurations 
by re-setting the inner and outer radii for the various disk components or 
simply by setting $\Sigma_r = 0$ for the cleared regions 
\citep[e.g.,][]{jensen96a}.  One expected result from these SED models is that 
the submillimeter emission should be significantly diminished for systems with 
a projected semimajor axis ($a_p$) on the order of a few tens of AU, but 
essentially identical to single stars for small and large $a_p$.  Previous 
observations have indicated that $a_p \lesssim 50 - 100$\,AU binaries have less 
submillimeter emission than single stars or wider binaries 
\citep{jensen94,jensen96a,osterloh95}, with the important exception of 
spectroscopic binaries \citep[$a \lesssim 1$\,AU; 
e.g.,][]{mathieu95,mathieu97}.  

Table \ref{binaries} gives a list of multiple stars in the sample and their 
projected separations.  This information and the data in Table 
\ref{results_table} have been compiled in Figure \ref{semimajor}, which shows 
the 850\,$\mu$m flux density and disk mass as a function of projected semimajor 
axis.  To be consistent with previous work, spectroscopic binaries as well as 
Class I sources have been excluded in this figure and the analysis that 
follows.  Unresolved higher-order multiple star systems ($\ge 3$ stars) without 
resolved observations in the literature were assigned the same $F_{\nu}$ or 
$M_d$ value for all projected separations.  Notes on assigning values for a few 
other systems are provided in Table \ref{binaries}.  We utilize a variety of 
two-sample statistical tests that incorporate upper limits to determine the 
probabilities that the 850\,$\mu$m flux densities and disk masses in various 
binary subsamples are drawn from \emph{different} parent distributions.  The 
total sample is separated into categories based on projected semimajor axis, 
resulting in three groups: close binaries with separations less than some 
critical value, $a_c$; wide binaries with separations larger than $a_c$; and 
single stars.  

Table \ref{binary_prob} lists the ranges of probabilities that the various 
subsamples for $a_c = 50$\,AU and 100\,AU differ from a sequence of survival 
analysis statistical tests performed with the ASURV software: the logrank, Peto 
\& Peto, Peto \& Prentice, and Gehan tests \citep[see the detailed descriptions 
by][]{feigelson85}.  The same tests were also performed for Class II objects 
only.  The results in Figure \ref{semimajor} and Table \ref{binary_prob} 
confirm the earlier conclusions of \citet{jensen94,jensen96a}: submillimeter 
flux densities and disk masses are significantly lower in close binaries ($a_p 
\le 50 - 100$\,AU) than wider or isolated systems and wide binaries essentially 
have the same disk masses as single stars.  These differences are greatest for 
a critical semimajor axis $a_c = 100$\,AU.  The results for the total sample 
(i.e., when Class III binaries are included) generally exhibit lower 
probabilities than the subsample of only Class II objects in Table 
\ref{binary_prob}, with the exception of the close and wide binary populations 
with the $a_c = 100$\,AU cutoff criterion.  These differences are likely due to 
the evolutionary behavior of disks between the Class II and III stages (see \S 
4), rather than an environmental effect in the multiple system.  The exact 
probabilities for the various subsamples appear to be fairly sensitive to the 
assignment of flux densities or disk masses in unresolved higher-order multiple 
systems.  High resolution interferometric observations are needed to determine 
the relative submillimeter contributions of individual components in these 
systems.

In a statistical sense, the presence of a companion in the range of 
$\sim$1$-$100\,AU decreases the apparent disk mass(es) in the system, 
presumably due to enhanced accretion onto the stars and/or dispersal into the 
local ISM.  However, multiple star systems \emph{still contain disks}, as 
evidenced by the relatively high detection rate in the submillimeter, $66 \pm 
10$\% for multiple systems compared to $58 \pm 8$\% for single stars, as well 
as other inner disk signatures in the optical and infrared 
\citep[e.g.,][]{white01}.  High-resolution interferometric measurements of 
disks in multiple systems have revealed a number of important exceptions to the 
statistical analysis above.  For example, the GG Tau A and UZ Tau systems both 
have small projected separations but very large disk masses, the former in a 
circumbinary disk and the latter in a pair of disks with four stellar 
components \citep{koerner93a,dutrey94,jensen96}.  Moreover, single-dish 
continuum surveys may be missing signatures of outer disks in close multiple 
systems: a close binary ($a_p = 32$\,AU) in the SR 24 triple system in 
Ophiuchus was found to have a large circumbinary gas disk detected in CO line 
emission but not in the continuum due to its low mass \citep{andrews05}.  The 
question of whether the large fraction of young stars in multiple systems could 
eventually harbor planetary systems will remain unanswered until more detailed 
case studies with interferometers \citep[e.g.,][]{jensen03} can confirm the 
properties of their disks.

\section{Discussion}

A summary of representative numbers derived from this submillimeter survey of 
Taurus-Auriga is provided in Table \ref{summ}.  Listed are submillimeter 
detection fractions as well as median values and standard deviations of disk 
masses and submillimeter continuum slopes for the total sample and various 
subsamples of interest.  Of the complete sample of 153 YSOs, $61 \pm 6$\% were 
detected for at least one submillimeter frequency, with a median $M_d \approx 5 
\times 10^{-3}$\,M$_{\odot}$ and $\alpha \approx 2.0$.  Single and multiple 
star systems have essentially identical detection rates and similar continuum 
slopes.  However, as discussed in detail in the previous section, closer 
binaries have statistically lower disk masses than wider systems or single 
stars.  Although Figure \ref{halpha} demonstrates that there is no direct 
correlation between $M_d$ or $\alpha$ and the equivalent width or luminosity of 
the H$\alpha$ emission line, there is an obvious difference in the 
submillimeter detection fraction between WTTS and CTTS disks.  The very high 
submillimeter detection rate for CTTS disks ($91 \pm 11$\%) is consistent with 
\emph{all} CTTSs having disk masses greater than 
$\sim$10$^{-4}$\,M$_{\odot}$.  The bulk of the detected WTTS disks are 
clustered near $W$(H$\alpha$) = 10\,\AA: when the WTTS/CTTS division criterion 
is slightly relaxed, this result suggests that nearly all WTTSs are either 
diskless or have very low disk masses.  Therefore, the equivalent width of the 
H$\alpha$ emission line appears to be a fairly robust predictor of the presence 
of a ``massive" disk.

Spectral energy distribution classifications of objects in the sample were 
determined based on power-law fits from 2 to 60\,$\mu$m (when possible) with 
data from the literature \citep[][and references 
therein]{strom89,weaver92,kh95,hartmann05}.  We adopt the classification 
breakdown of \citet{greene94}, using the values of the power-law index $n$ 
(defined by $\nu F_{\nu} \propto \nu^n$) to distinguish between Class I, Flat 
Spectrum, Class II, and Class III sources.  The derived classifications are 
listed in Table \ref{results_table}.  Although it is not absolutely calibrated 
in time, the YSO evolution sequence defined by the shape of the infrared SED is 
certainly indicative of changes in the physical structure of the inner regions 
of the circumstellar disk and/or envelope.  With the large sample of 
submillimeter data presented above, we can address the issue of corresponding 
changes in the physical properties of the outer disk.  

Motivated by the differences in the detection rates and median properties of 
the various SED classes listed in Table \ref{summ}, the same survival analysis 
two-sample statistical tests used in \S 3.4 were employed to determine the 
probabilities that the 850\,$\mu$m flux densities, inferred disk masses, and 
submillimeter continuum slopes for various SED and H$\alpha$ line strength 
classes are drawn from different parent populations.  The test results are 
given in Table \ref{evol_prob}, and the cumulative distributions of $F_{\nu}$, 
$M_d$, and $\alpha$ for different classes are shown in Figure \ref{diskevol}.  
There are statistically significant progressions of decreasing submillimeter 
flux densities, disk masses, and continuum slopes along the infrared SED 
evolution sequence.  Flat Spectrum objects fit between Class I and II objects 
in these respects, with somewhat more similarity to the latter.  Incorporating 
the Flat Spectrum objects with either the Class I or II objects does not make 
any significant difference in these results.  Apparently the properties of the 
outer disk/envelope evolve along a similar evolutionary sequence as the inner 
disk.

Figure \ref{diskevol} clearly shows that Class I objects have significantly 
larger submillimeter flux densities, disk masses, and continuum slopes than 
Class II objects.  It should again be stressed that the extent to which these 
values are representative of Class I \emph{disks}, rather than disks + inner 
envelopes, is questionable.  It has been suggested that many Class I disk 
properties could be mimicked by a Class II disk viewed at high inclination 
\citep[e.g.,][]{chiang99,white04}.  It is shown in Figure \ref{incl} that for a 
given mass, a high inclination angle produces both a lower flux density and a 
lower continuum slope; the opposite is seen in Figure \ref{diskevol} and Table 
\ref{evol_prob}.  If the Class I emission is primarily from a disk (with only a 
comparatively small contribution from the inner envelope), a simple 
re-orientation of a Class II disk will not reproduce the Class I 
\emph{submillimeter} properties without additional changes in mass, 
temperature, or opacity.  The distributions found here of the \emph{empirical} 
(model-independent) flux densities and continuum slopes corroborate the 
original picture of Class I sources as disk + envelope systems: the higher flux 
densities and ``disk" masses may be due to additional envelope mass, and the 
higher continuum slopes may be due to the less-processed (i.e., lower amount of 
grain growth) dust in the envelope.  A large interferometric sample will be 
required to definitively settle the issues involved in a comparison of Class I 
and II disks.

Unfortunately, there are no measurements of a submillimeter continuum slope for 
any of the Class III objects in the sample, and the 3-$\sigma$ upper limits are 
too large to make any definitive statements on an evolutionary trend in 
$\alpha$.  The direct relationships between $M_d$ or $\alpha$ and the infrared 
SED slope are shown in Figure \ref{irexc_smm}.  There is no \emph{direct} 
correlation with $M_d$,\footnote{A significant correlation exists when L1551 
IRS 5 and L1551 NE are included (the two points in the upper left corner), but 
these objects likely have large contributions to $M_d$ from their envelopes, 
and so have been excluded in this part of the analysis.} but differences 
between the SED classes in general are apparent.  A more steady decrease in 
$\alpha$ is seen across the evolution sequence, which if it continues would 
imply very shallow continuum slopes for Class III disks ($\alpha \sim 1 - 
1.5$).  The Spearman rank correlation coefficient in this case is $-0.50$, with 
a 99.98\% confidence level (3.7-$\sigma$).  The best-fit linear relation 
between the submillimeter and infrared continuum slopes is $\alpha = 0.40(\pm 
0.04) n + 2.09(\pm 0.03)$.  This trend can not be explained solely by 
decreasing optical depths in the disks along an evolutionary sequence: lower 
optical depths produce steeper continuum slopes.  Another effect, such as a 
shallow opacity function or temperature/surface density evolution, must be 
acting to decrease $\alpha$ in this manner.  However, interferometric 
observations of the Class I sources at several wavelengths would be required to 
confirm the validity of this trend.

The submillimeter detection fraction and the fraction of objects with a 
near-infrared ($K_s-L$) excess are identical: $60 \pm 7$\%.\footnote{The values 
given here and in Table \ref{summ} are slightly different because a small 
fraction of the objects in the sample do not have $L$-band measurements in the 
literature.}  Figure \ref{ccd} is a near-infrared color-color diagram that 
indicates the sources with submillimeter detections.  Of the 6 sources with 
infrared excesses that were not detected in the submillimeter, 5 could have 
anomalous colors due to mismatched photometry and/or infrared companions (see 
the Appendix).  Three of the 55 sources with essentially no near-infrared 
excess, or $5 \pm 3$\%, were detected in the submillimeter: GM Aur, V836 Tau, 
and CoKu Tau/4.\footnote{A fourth source, BP Tau, appears to fit in this 
category in Figure \ref{ccd}.  However, this is likely due to mismatched 
photometry from 2MASS and the literature, because BP Tau has a clear excess in 
homogeneous datasets, e.g. \citet{kh95}.}  All three of these YSOs also have 
mid- and far-infrared emission, indicating that the lack of near-infrared 
excess may be due to a clearing of dust in the inner $\sim$1\,AU of their 
disks.  In addition to these ``transition" objects, three Class III sources, 
V807 Tau, FW Tau, and LkH$\alpha$ 332/G1, were also detected in the 
submillimeter (a $5.6 \pm 3.2$\% detection rate), along with another possible 
Class III candidate whose SED classification remains to be confirmed due to 
lack of infrared data (HQ Tau).  However, in general a YSO with a near-infrared 
excess also has submillimeter emission consistent with a disk mass greater than 
$\sim$10$^{-4}$\,M$_{\odot}$, and vice versa.  The small fraction of objects, 
less than 10\%, with evidence for an outer disk (from submillimeter data) and 
no inner disk suggests that the timescale for the disappearance of \emph{both} 
infrared and submillimeter disk emission is relatively short; no more than a 
few hundred thousand years (i.e., $\lesssim 10$\% of the typical YSO age in 
Taurus-Auriga).  In agreement with the comparatively low detection fraction for 
WTTS disks ($16 \pm 5$\%) and other similar analyses 
\citep[e.g.,][]{skrutskie90,wolk96,duvert00}, these results imply that the 
inner and outer disk dissipate, or become unobservable, almost simultaneously.

The physical mechanism responsible for the rapid and essentially radially 
constant ``disappearance" timescale remains to be explained.  Viscous accretion 
onto the central star alone does not readily produce the apparently rapid 
inner-outer disk dissipation \citep{hollenbach00}.  In fact, evolution under 
accretion processes predicts only small changes in submillimeter emission with 
time \citep[e.g.,][]{hartmann98}.  Models which incorporate the ultraviolet 
photoevaporation of the outer disk along with viscous accretion have 
more success in reproducing the inferred dissipation timescale, particularly 
for disk emission out to $\sim$100\,$\mu$m \citep{clarke01,armitage03}.  
However, these ``ultraviolet switch" models also suggest that submillimeter 
emission is relatively unaffected, and could therefore predict a fairly large 
fraction of WTTSs or Class III sources with submillimeter emission.  
\citet{clarke01} suggest that the low observed fraction of such transition 
objects noted by \citet{duvert00} and confirmed by the larger sample presented 
here may be accomodated in their models if different surface density profiles 
or viscosity values are adopted.

An alternative explanation to actually \emph{losing} disk material, onto the 
star or elsewhere, is a process which renders the dust invisible to 
conventional observations.  A compelling possibility is the collisional 
agglomeration of dust grains in the disk.  Accelerated by gravitational 
settling to the disk midplane, the characteristic grain growth timescales even 
at fairly large disk radii are thought to be shorter than the transition 
timescale inferred above \citep[e.g.,][]{weidenschilling93}.  Perhaps the grain 
growth process has rendered the disks around many of the evolved (e.g., Class 
III) sources invisible by creating a significant population of large 
($\sim$cm-sized) grains which are inefficient emitters at both infrared and 
submillimeter wavelengths.  If this is to be the case, any collisional 
fragmentation process of the aggregate grains should not produce more than 
$\sim$10$^{-4}$\,M$_{\odot}$ of particles which are efficient submillimeter 
emitters.  The shallow submillimeter slopes measured in \S 3.3 and the implied 
low values of the opacity index $\beta$ lend some credibility to the grain 
growth argument.  Theoretical studies indicate that $\beta$ values such as 
those inferred in this sample ($\beta \sim 1$ or less) can be the result of a 
significant population of large grains \citep{miyake93,pollack94,dullemond05}.  
The collisional growth of dust grains has also been inferred from submillimeter 
observations of both young \citep[e.g.,][]{beckwith91,mannings94,koerner95b} 
and old \citep[e.g.,][]{calvet02,hogerheijde02} low mass disks, and 
particularly for those around the more massive Herbig Ae stars 
\citep{testi01,testi03,natta04}.  Complementary studies of scattered light 
\citep[e.g.,][]{mccabe03,duchene04b} and mid-infrared spectra 
\citep{meeus03,przygodda03,vanboekel04,kessler05} also suggest that typical 
grain sizes are larger in these disks than for the ISM.  The feasibility of 
this hypothesis depends critically on coupling with another mechanism which can 
diminish the accretion of \emph{gas}, and therefore also explain the low 
submillimeter detection fraction for WTTSs.

The distribution of $M_d$ shown in Figure \ref{F850_CDF} indicates that typical 
disks have masses significantly lower than those required by two of the leading 
theoretical models for giant planet formation.  Both the core accretion 
\citep{pollack96} and disk instability \citep{boss98} scenarios require disk 
masses at least a few times that of the MMSN to form a Jupiter-like planet; 
roughly an order of magnitude higher than the median mass inferred for 
Taurus-Auriga disks.  Radial velocity surveys suggest that roughly 10\% of 
stars harbor a gas giant planet within a few AU \citep[e.g.,][]{marcy05}, with 
the prospect that better sensitivity to long-period planets could significantly 
increase that fraction \citep[e.g.,][]{fischer01}.  This shows that planet 
formation is a fairly common process.  In order for that to be the case, the 
disk mass distribution constructed in \S 3.2 needs to be reconciled with the 
theoretical requirements of the planet formation models.  Possible remedies 
could be extracted from changes to the simple disk model used in \S 3: for 
example, adjustments to the disk surface density profile, or a significant 
decrease in the normalization of the opacity function.  Unfortunately, 
solutions like these will remain untested until more advanced observations 
become available (e.g., interferometers with $\sim$0\farcs1 spatial 
resolution).  A likely alternative explanation, as discussed above, is that a 
significant fraction of the disk mass is locked up in large grains or 
planetesimals which are inefficient emitters at submillimeter wavelengths.  

\section{Summary}

We have conducted a sensitive, multiwavelength submillimeter survey of 153 
Taurus-Auriga YSOs in an effort to analyze properties of the outer regions of 
circumstellar dust disks.  Some of the key results from this survey are 
summarized here:
\begin{itemize}
\item The disk mass (or submillimeter flux density) distribution function is 
well matched with a log-normal distribution centered around $5 \times 
10^{-3}$\,M$_{\odot}$ with a large dispersion (0.5\,dex).  The vast majority 
of disks in Taurus-Auriga have substantially lower masses than is thought to be 
required for giant planet formation.  However, a significant fraction of the 
disk mass could be stored in large grains or planetesimals which do not 
contribute to the submillimeter emission.
\item We provide the largest set of submillimeter continuum slope measurements 
of YSOs to date.  The empirical behavior of the continuum from 350\,$\mu$m to 
1.3\,mm is well-described by $F_{\nu} \propto \nu^{2.0 \pm 0.5}$, which is much 
flatter than for the interstellar medium.  The low observed slope values are 
probably due to a combination of optical depth effects \citep{beckwith91} and 
an inherently shallow opacity function from the top-heavy grain size 
distribution produced by collisional agglomeration of material in the disk.
\item There do not appear to be any links between stellar and disk properties 
in the sample, although the stellar masses and ages span a relatively limited 
range.  The median disk to star mass ratio is $\sim$0.5\%.
\item Submillimeter flux densities and disk masses are statistically lower for 
stars with close companions (projected semimajor axes less than $\sim$100\,AU) 
than for wider binaries or single stars.  However, multiple star systems often 
still \emph{contain} disks, regardless of their projected separations.  
Multiple star systems with wider separations have flux densities and disk 
masses comparable to single stars.  
\item In general, the standard signatures of the inner disk (e.g., accretion 
diagnostics or infrared excess emission) are accurate predictors of a disk 
mass greater than $\sim$10$^{-4}$\,M$_{\odot}$.
\item Statistically significant changes in the distribution functions of 
submillimeter flux densities, disk masses, and continuum slopes are found for 
the progressive stages of YSO evolution inferred from inner disk observations.  
These measured outer disk properties \emph{decrease} from Class I $\rightarrow$ 
II $\rightarrow$ III objects as well as for CTTSs $\rightarrow$ WTTSs.  The 
implication is that the inner and outer disk develop along a similar 
evolutionary sequence.  A multiwavelength interferometric survey of Class I 
objects would be very useful for determining the relative contributions of a 
disk and inner envelope in these systems for a more sophisticated comparison 
with their presumably more evolved counterparts (Class II and III objects).
\item Only a small fraction of objects ($< 10$\%) which have no inner disk 
signatures were detected in the submillimeter, suggesting that \emph{both} 
infrared and submillimeter disk emission disappear on a similar timescale 
(within $\sim$10$^5$ years of each other).  There are two timescales in 
operation for disk evolution: (1) the relatively long ($\sim$5 to 10\,Myr) 
lifetime of Class II/CTTS disks, and (2) the rapid (a few $\times$ $10^5$ 
years) transition period to Class III/WTTS disks.  Understanding the mechanisms 
responsible for these timescales, particularly the trigger for the transition 
stage, remain key problems in disk evolution.  Some possible explanations for 
the essentially radially-independent disk dissipation timescale include viscous 
accretion with photoevaporation by the central star \citep[e.g.,][]{clarke01} 
or rapid grain growth in the early stages of planet formation 
\citep[e.g.,][]{weidenschilling93}.  
\end{itemize}

\acknowledgments
We acknowledge useful conversations, suggestions, and advice from Michael Liu, 
Lee Hartmann, Ted Simon, and Alan Boss.  We would like to thank the JCMT and 
CSO support staffs, and in particular Colin Borys, for their assistance.  An 
anonymous referee provided valuable criticism which helped improve this paper.  
This work was supported by NSF grant AST-0324328.  This research has made use 
of the JCMT archive at the Canadian Astronomy Data Center, which is operated by 
the Dominion Astrophysical Observatory for the National Research Council of 
Canada's Herzberg Institute of Astrophysics and the NASA/IPAC Infrared Science 
Archive, which is operated by the Jet Propulsion Laboratory, California 
Institute of Technology, under contract with the National Aeronautics and Space 
Administration.  

\appendix

\section{Comments on Disk Models}

A simple disk structure model was presented in \S 3.1 and used to generate SEDs 
of thermal dust emission and derive a relationship between the submillimeter 
flux density and disk mass.  A lack of SED data severely limits the number of 
degrees of freedom in this modeling, and therefore a number of parameters were 
fixed: inclination, inner and outer radii, and the power law indices of the 
radial surface density and frequency spectrum of the opacity.  Here we 
determine the effects that reasonable alternative values for these fixed 
parameters would have on the $F_{\nu} - M_d$ relationship.  Readers would also 
benefit from the mathematical formalism discussed in detail by \citet{bscg90}.  
The analysis is done comparatively, relative to a fiducial disk model based on 
the median values discussed in the text: $i = 0$\degr, $p = 1.5$, $T_1 = 
150$\,K, $q = 0.6$, $r_{\circ} = 0.01$\,AU, $R_d = 100$\,AU, and $\beta = 1$.  
A set of 850\,$\mu$m flux densities was computed using Equation 2 for a grid of 
disk masses with this fiducial parameter set excepting one of the previously 
fixed variables.  The flux density $-$ disk mass relationships for various 
values of $p$, $q$, $r_{\circ}$, $R_d$, $T_1$, and $\beta$ are shown in Figure 
\ref{FMparams}.  It should be noted that these plots are only intended to 
illustrate the effects of changing a single parameter in these models.  In 
reality, the parameters are usually somewhat coupled, and therefore changes in 
one parameter affect others: such coupled effects are not considered in this 
simple comparative analysis.

Larger values of $q$, $r_{\circ}$, $R_d$, and $\beta$ all result in a larger 
disk mass for a given flux density.  The middle panels in Figure \ref{FMparams} 
indicate that the disk boundaries play only a minor role in setting the 
$F_{\nu} - M_d$ relationship.  The roles of the opacity index and temperature 
normalization are roughly those expected from the optically thin assumption 
given in Equation 1, where $F_{\nu} \propto \kappa_{\nu} M_d T^{-1}$.  For 
low-mass disks (i.e., $M_d \lesssim 10^{-2}$\,M$_{\odot}$), the shape of the 
radial surface density profile can have a significant effect on the 
submillimeter flux density because most of that emission is optically thin.  
For the same reason, the radial temperature behavior dictated by $q$ has an 
opposite effect and impacts the flux densities for the more massive, 
optically thick disks.  The largest relative deviations in the flux density $-$ 
disk mass relationship are set by the parameters which describe the radial 
temperature profile; $T_1$ and $q$.  Fortunately, the parameters of the 
temperature profile can be reliably determined for individual disks using 
observations in the mid- and far-infrared.  Sensitive observations with the 
\emph{Spitzer} Space Telescope will essentially ensure that the opacity is the 
single dominant uncertainty in the determination of a disk mass from 
submillimeter observations.

In Figure \ref{aMparams}, the same procedure as above was used to examine the 
effects of different fixed parameter choices on the relationship between the 
submillimeter continuum slope and the disk mass (note that here we have fixed 
$\beta = 1$ as a reference value).  Once again, the disk boundaries play a 
negligible role in the relationship.  The radial surface density index has only 
a small impact on the $\alpha - M_d$ relationship in general, although a 
constant surface density disk ($p = 0$) gives a roughly constant submillimeter 
continuum slope (and lower than for larger values of $p$) for low-mass disks.  
The parameters of the temperature profile again show the largest deviations 
from the fiducial $\alpha - M_d$ relationship.  The effects of a changing 
$\beta$ in this relationship are also large.  See \citet{beckwith91} and 
\citet{mannings94b} for a mathematical description of what is shown in these 
plots.

An alternative view of the relationship between the observed continuum slope 
($\alpha$) and the power law index of the frequency behavior of the opacity 
($\beta$) can be obtained directly from these models.  Figure \ref{ab} shows 
computed continuum slopes as a function of $\beta$ for various disk masses.  
The relationship is essentially linear for $M_d \lesssim 0.1$\,M$_{\odot}$ with 
slopes of nearly unity independent of disk mass, but significantly different 
intercept values.  All of the curves fall below the nominal $\beta = \alpha - 
2$ line which is representative of optically thin emission in the 
Rayleigh-Jeans limit.  This effect is due to the failure of the Rayleigh-Jeans 
limit and the fraction of optically thick submillimeter emission, which can be 
fairly high at the shortest wavelengths \citep{bscg90,beckwith91,mannings94b}.  
This synthetic grid illustrates the conclusion in \S 3.3 that the roughly 
gaussian distribution of $\alpha$ centered around $\sim$2 suggests that the 
opacity index is likely between $\sim$0.5 and 1.5, and certainly less than the 
ISM value of 2 unless the disk masses are severely underestimated.  

Figure \ref{tau} was generated from the fiducial model (the top panel also 
assumes the median $M_d = 0.005$\,M$_{\odot}$) described above to illustrate 
that the assumption of optically thin emission in the submillimeter continuum 
is not always valid \citep[as pointed out by][]{bscg90}.  The disk becomes 
optically thin at the radius, $r_1$, where $\tau_{\nu} = \kappa_{\nu} 
\Sigma_{r_1} = 1$.  That criterion can be solved to determine $r_1$ using 
Equation 4 and the relationship between $\Sigma_{\circ}$ and $M_d$, giving
\begin{equation}
r_1 = \left(\kappa_{\nu} \Sigma_{\circ} \right)^{1/p} r_{\circ} = \left[ \frac{(2-p) \kappa_{\nu} M_d}{2\pi (R_d^{2-p} - r_{\circ}^{2-p})} \right]^{1/p},
\end{equation}
when $p \ne 2$; inserting the parameters fixed in \S 3.1 gives $r_1 \approx 184 
(M_d/\rm{M}_{\odot})^{2/3}$\,AU.  The fraction of the submillimeter flux 
density from optically thick emission, $\Delta$ \citep[defined as the ratio of 
the flux density from $r < r_1$ to the total flux density: see][]{bscg90} 
increases exponentially until $\Delta = 1$ around 0.4 to 1\,Jy, depending on 
the wavelength.  Even faint submillimeter sources have $\sim$15\% of their flux 
densities generated in the innermost (radially), optically thick regions of the 
disk.  In terms of the fraction of the disk mass which gives rise to optically 
thick emission (the ratio of the integrated surface density from $r_{\circ}$ to 
$r_1$ to the total disk mass), a more gradual trend with disk mass is 
present.  Roughly 25\% of the mass in a MMSN disk with the fiducial parameter 
set contributes optically thick emission in the submillimeter.  Comparison of 
the right panel in Figure \ref{tau} with the information in Figures 
\ref{Md_F850} and \ref{ab_Md} demonstrate the effects these relatively high 
optical depths have on estimating $M_d$ and $\alpha$ (and subsequently $\beta$) 
from observations.  

\section{Comments on Individual Sources}

There are six objects in the sample with apparent near-infrared excesses, but 
no detected submillimeter emission: DP Tau, JH 223, CoKu Tau/3, FV Tau/c, CZ 
Tau, and CIDA-3.  Three of these six (CoKu Tau/3, FV Tau/c, and CZ Tau) are 
multiple stars (see Table \ref{binaries}), introducing the possibility that the 
near-infrared excess is exaggerated by a red companion.  The infrared colors of 
JH 223, CoKu Tau/3, CZ Tau, and CIDA-3 also seem questionable in light of the 
fact that these sources are WTTSs.  Therefore, multiplicity and/or mismatched 
photometry from the literature for WTTSs could explain why submillimeter 
emission was not detected for most of these sources.  The non-detection of DP 
Tau is more puzzling.

As mentioned in \S 4, there are three sources without near-infrared excesses 
which are detected in the submillimeter (and mid- and far-infrared): GM Aur, 
V836 Tau, and CoKu Tau/4.  Supplementary to these transition objects are the 
Class III sources detected at 850\,$\mu$m: V807 Tau, FW Tau, and LkH$\alpha$ 
332/G1.  These objects have presumably evacuated the inner portions of their 
disks, perhaps due to clearing from the planet formation process \citep[see, 
e.g.,][regarding CoKu Tau/4]{forrest04}.  Examinations of the SEDs for this 
sample highlight some other interesting sources in terms of possible 
near-infrared deficits: FM Tau, FQ Tau, UX Tau, and HK Tau.  Seven late-type 
(excluding V892 Tau and SU Aur) WTTS disks are detected in the submillimeter: 
V773 Tau, IQ Tau, UX Tau, IT Tau, CoKu Tau/4, LkH$\alpha$ 332/G1, and V836 
Tau.  A number of other sources detected in the submillimeter but with very 
little information between $\sim$2 and 850\,$\mu$m include FY Tau, GN Tau, 
CIDA-7, CIDA-8, CIDA-9, and HQ Tau.  These are prime targets for mid- and 
far-infrared observations from both the ground and the \emph{Spitzer} Space 
Telescope.  One final source worth further investigation is CY Tau.  This 
object has an unique SED, well-described by a shallow power law from 
$\sim$2\,$\mu$m out to at least 1.3\,mm.  The models in \S 4 are clearly 
inappropriate for this case, as they predict very large disk masses ($M_d \sim 
10$\,M$_{\odot}$).  Updated mid- and far-infrared photometry and spectroscopy 
of this source, along with resolved observations of the dust content may clear 
up the true nature of the circumstellar material.

\clearpage

\clearpage

\clearpage

\clearpage

\clearpage

\clearpage

\clearpage

\clearpage

\begin{figure}
\epsscale{0.55}
\plotone{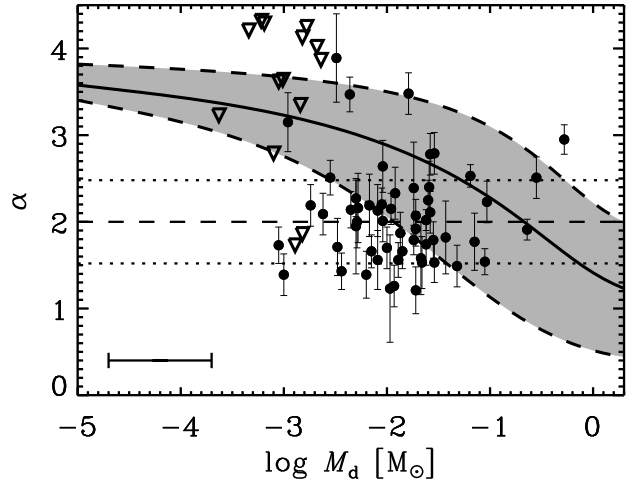}
\figcaption{The submillimeter continuum slope values ($\alpha$: defined as 
$F_{\nu} \propto \nu^{\alpha}$) plotted against 
the logarithm of the disk mass.  The open triangles are 3-$\sigma$ upper limits 
on $\alpha$.  A typical error bar on the disk mass which incorporates the 
systematic uncertainties ($\sim$0.5\,dex) due to the unknown value of the 
opacity index is shown in the lower left corner.  The dashed horizontal line 
marks the best-fit mean value of $\alpha$, determined from the cumulative 
distribution shown in Figure \ref{alpha_CDF}.  The dotted lines parallel to it 
mark the 1-$\sigma$ standard deviation from the mean.  The thick, solid curve 
marks the behavior of $\alpha(M_d)$ for the median disk model with $\beta = 
2$.  The shaded region marks the $\beta = 2$ curves for extreme temperature 
profiles: the lower boundary for $T_1 = 75$\,K and $q = 0.75$ and the upper 
boundary for $T_1 = 300$\,K and $q = 0.35$.  Much of the sample appears to have 
$\beta < 2$, indicating a change in the opacity function in disk material 
relative to the interstellar medium which could be a signature of the 
collisional growth of dust grains. \label{ab_Md}}
\end{figure}   

\clearpage

\begin{figure}
\epsscale{1.0}
\plotone{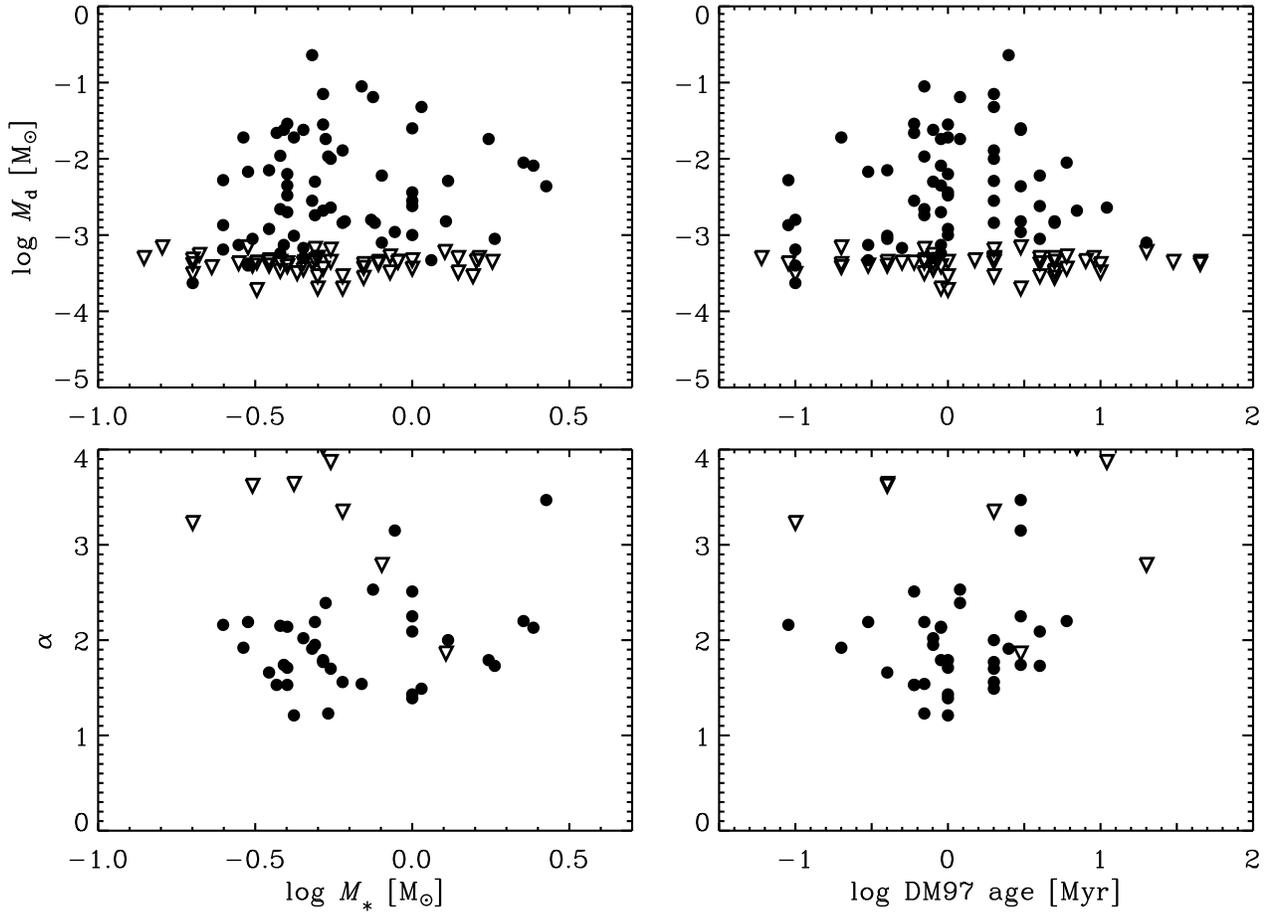}
\figcaption{Plots showing the relationships between disk masses or 
submillimeter continuum slopes and stellar masses or ages.  Filled circles are 
detections and open triangles are 3-$\sigma$ upper limits. \label{starscat}}
\end{figure}

\clearpage

\begin{figure}
\epsscale{0.55}
\plotone{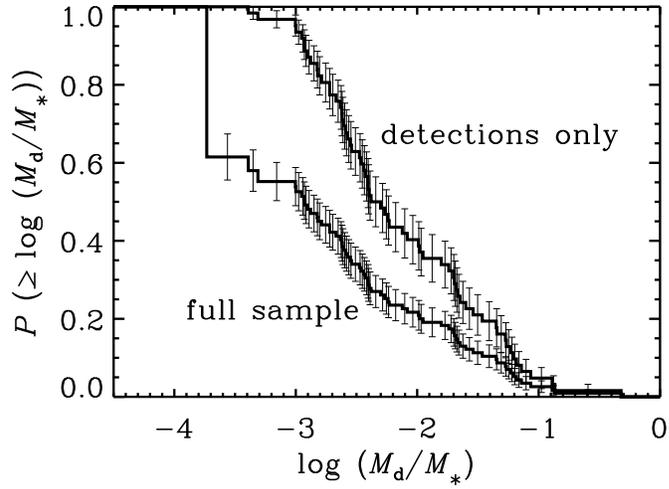}
\figcaption{The cumulative distributions of the disk to star mass ratio for 116 
objects in the full sample (constructed with the Kaplan-Meier estimator) and 61 
objects which were detected at submillimeter wavelengths.  The ordinate values 
represent the probability that an object in the sample has a mass ratio greater 
than or equal to the abscissa value.  Less than 10\% of the sample has a mass 
ratio which could result in a gravitational disk instability, and even this 
small fraction may be contaminated by mass in an envelope.  The median disk to 
star mass ratio is 0.5\%.  \label{CDF_MDMS}}
\end{figure}

\clearpage

\begin{figure}
\epsscale{0.5}
\plotone{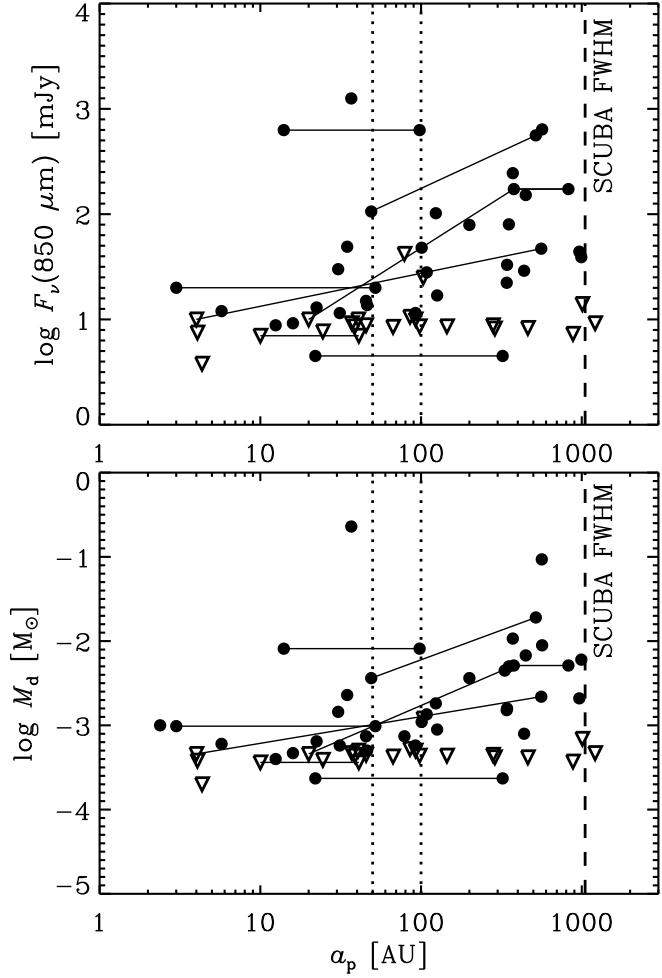}
\figcaption{The relationship between the 850\,$\mu$m flux density (\emph{top}) 
or disk mass (\emph{bottom}) and projected semimajor axis for multiple star 
systems which were not resolved by the observations.  Open triangles are 
3-$\sigma$ upper limits, and higher order systems are connected with solid 
lines.  The dashed vertical lines mark the FWHM beam radius of SCUBA at 
850\,$\mu$m, while the dotted lines mark 50 and 100\,AU.  For Class II multiple 
systems, the flux densities and disk masses in close binary systems ($a_p \le 
50$ or 100\,AU) are statistically lower than those in wider binary systems or 
single stars (see Table \ref{binary_prob}). \label{semimajor}}
\end{figure}

\clearpage

\begin{figure}
\epsscale{1.0}
\plotone{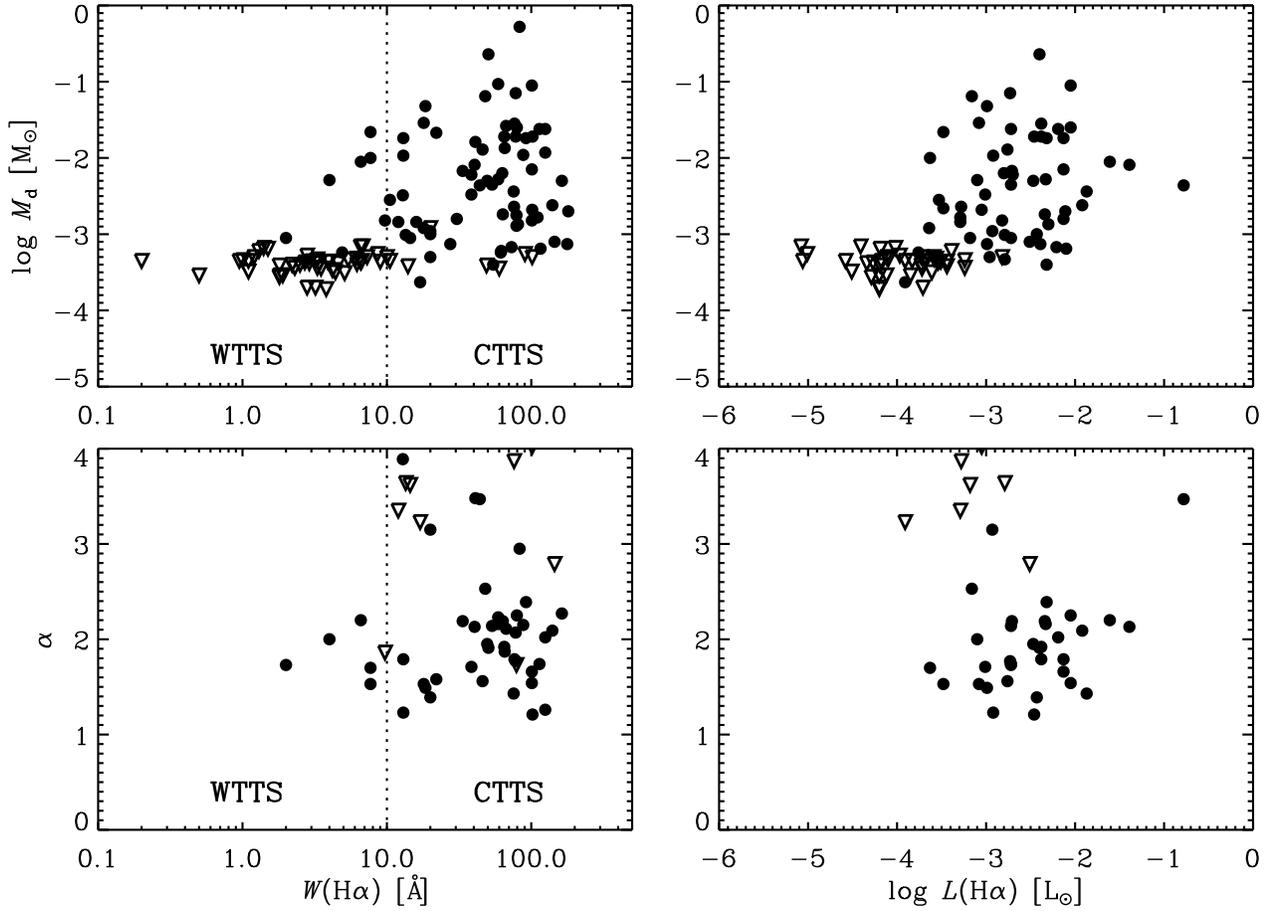}
\figcaption{The behavior of disk mass or submillimeter continuum slope for 
various ranges of the equivalent width or luminosity of the H$\alpha$ emission 
line.  Filled circles are detections and open triangles are 3-$\sigma$ upper 
limits.  The dotted vertical lines mark the boundary between WTTSs and CTTSs.  
Luminosities in the H$\alpha$ line were computed from equivalent widths and a 
linear relationship between the local continuum flux and the 
extinction-corrected $R$-band flux \citep[][see their Figure 1a]{reid95}. 
\label{halpha}}
\end{figure}

\clearpage

\begin{figure}
\epsscale{0.9}
\plotone{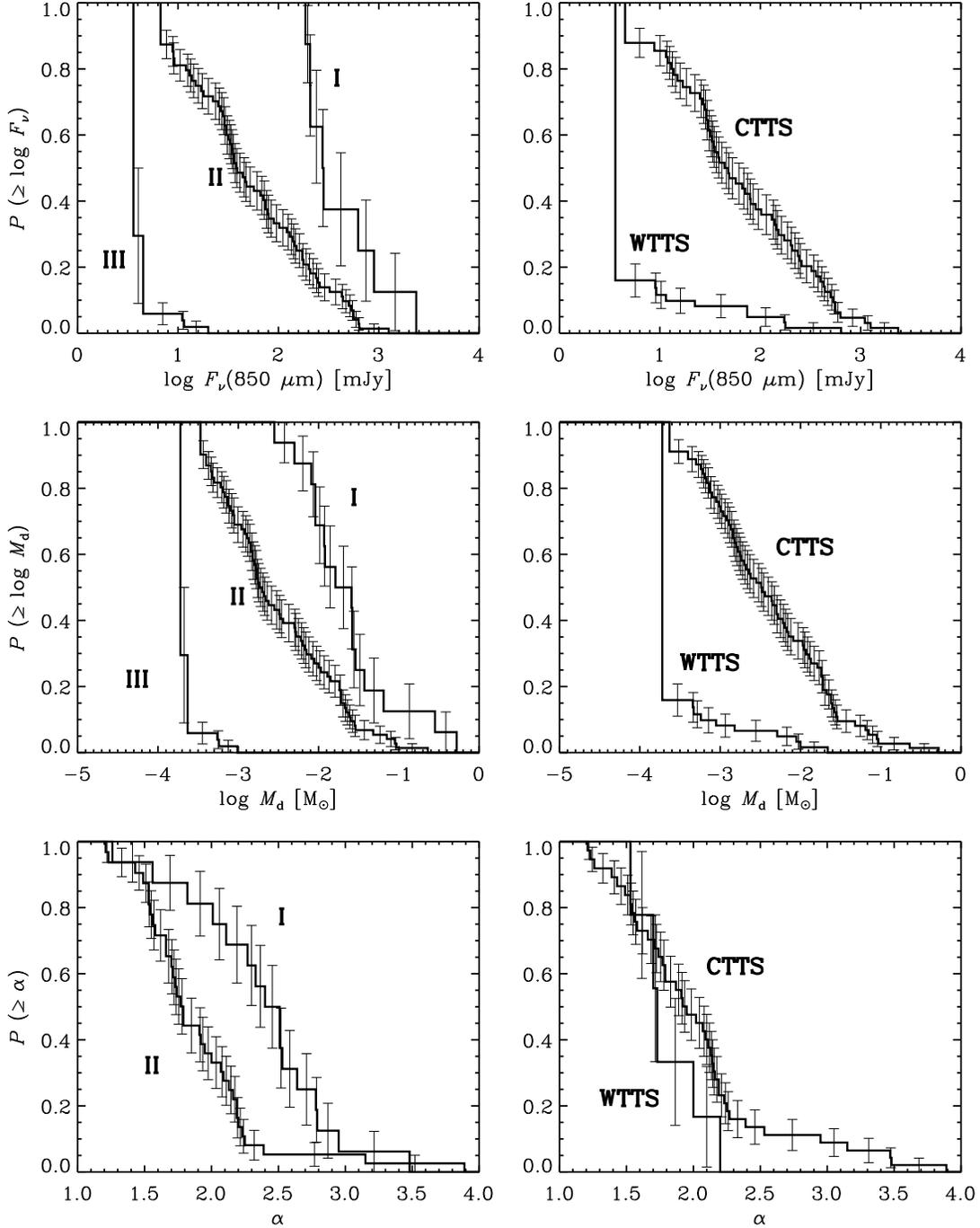}
\figcaption{Evidence for outer disk evolution from the distribution functions 
of the 850\,$\mu$m flux densities (\emph{top}), circumstellar disk masses 
(\emph{middle}) and submillimeter continuum slopes (\emph{bottom}).  Each 
cumulative distribution is labeled with the infrared SED or $W$(H$\alpha$) 
classification from which it was constructed.  Statistical tests (see Table 
\ref{evol_prob}) confirm that there is a decrease in $F_{\nu}$ and $M_d$ from 
Class I $\rightarrow$ II $\rightarrow$ III and CTTS $\rightarrow$ WTTS sources, 
and that $\alpha$ is also diminished in the evolution from Class I 
$\rightarrow$ II sources.  \label{diskevol}}
\end{figure}

\clearpage

\begin{figure}
\epsscale{0.5}
\plotone{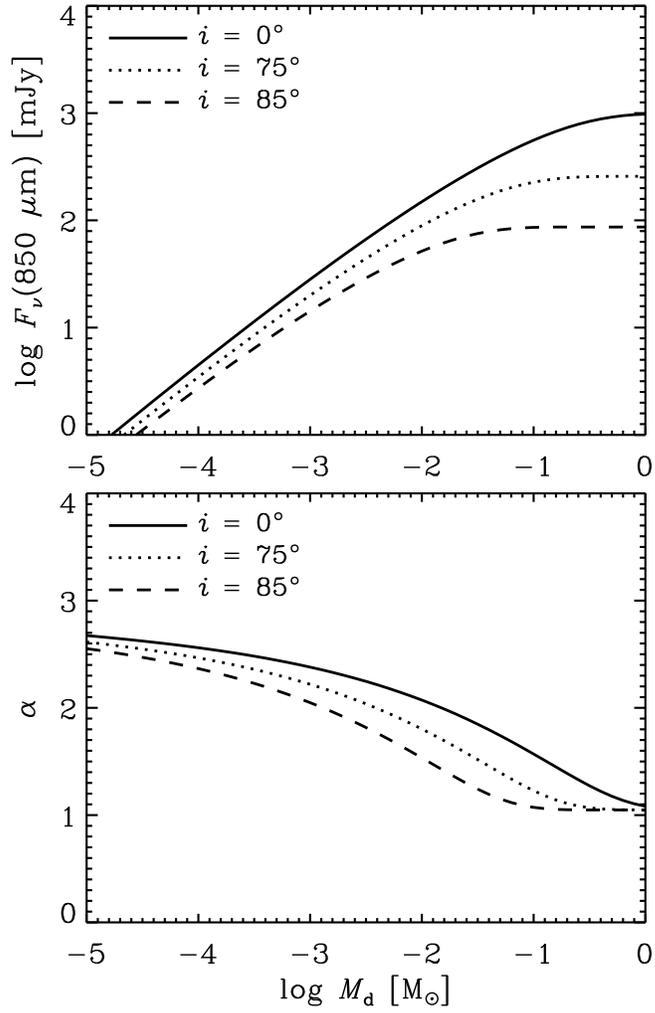}
\figcaption{Illustration of the effect on the relationships between the disk 
mass and 850\,$\mu$m flux density or submillimeter continuum slope introduced 
by increasing the inclination angle of a fiducial circumstellar disk (the 
median disk model described in the text).  Higher inclination angles (closer to 
edge-on disks) produce lower flux densities and continuum slopes for a given 
disk mass.  \label{incl}}
\end{figure}

\begin{figure}
\epsscale{0.5}
\plotone{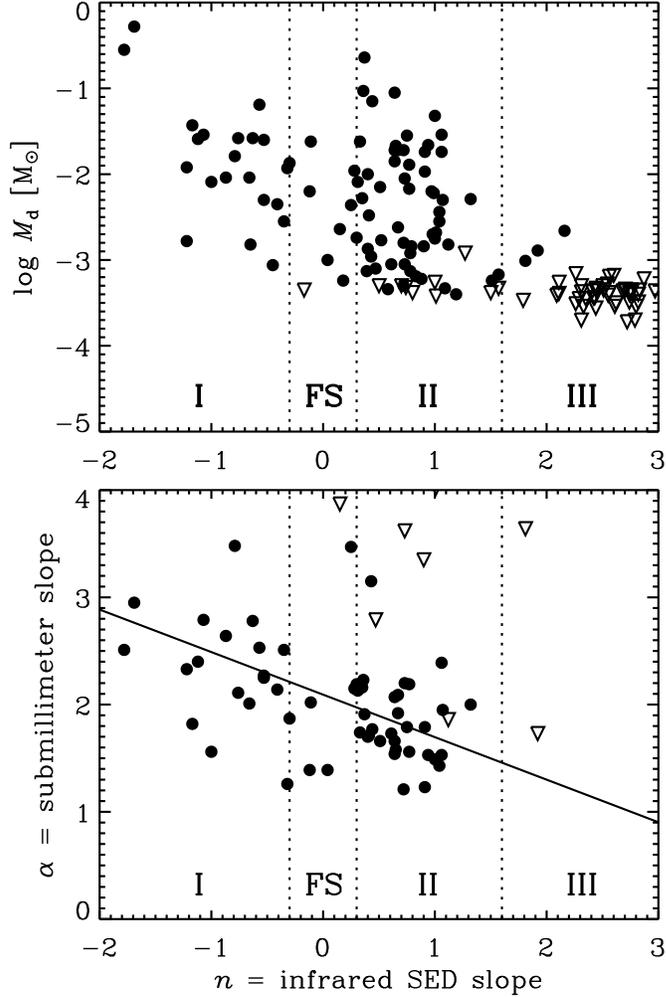}
\figcaption{Relationship between the disk mass (\emph{top}) or submillimeter 
continuum slope (\emph{bottom}: $\alpha$ defined by $F_{\nu} \propto 
\nu^{\alpha}$) and the slope of the infrared SED from $\sim$2 to 60\,$\mu$m, 
defined as the index $n$ such that $\nu F_{\nu} \propto \nu^n$.  The vertical 
dotted lines mark the borders between various SED classifications.  Filled 
circles are detections and open triangles are 3-$\sigma$ upper limits.  There 
is no direct correlation between $n$ and $\log{M_d}$.  A correlation 
(3.7-$\sigma$ with Spearman rank coefficient of $-0.50$) is seen between $n$ 
and $\alpha$, where the best-fit linear relationship has been overlaid as a 
solid line: $\alpha = -0.40(\pm0.04)n + 2.09(\pm0.03)$.  The rms residual 
dispersion around the best-fit line is 0.24, which is roughly the 1-$\sigma$ 
error on $\alpha$ expected from absolute flux calibration uncertainties.  
\label{irexc_smm}}
\end{figure}

\clearpage

\begin{figure}
\epsscale{0.55}
\plotone{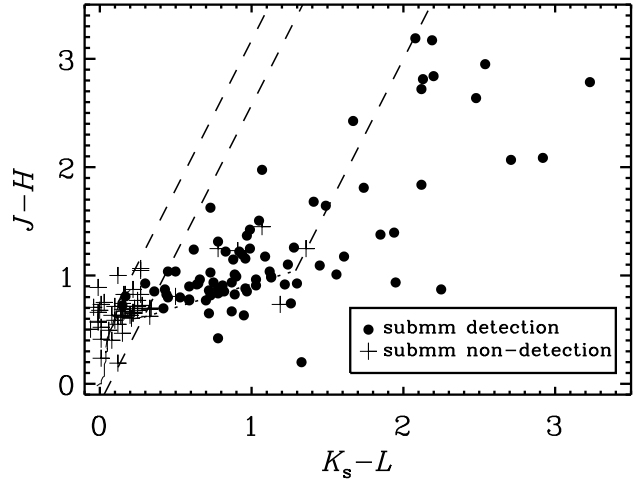}
\figcaption{A near-infrared color-color diagram of the sample constructed from 
2MASS data and $L$-band data from the literature.  The solid curves mark the 
intrinsic colors of dwarf and giant stars.  Dashed lines denote the reddening 
vectors based on the extinction law of \citet{cohen81}, converted to the 2MASS 
photometric system.  The dotted line corresponds to the classical T Tauri star 
locus derived by \citet{meyer97}.  Filled circles are sources which are 
detected for at least 1 submillimeter wavelength (between 350\,$\mu$m and 
1.3\,mm), and crosses are undetected.  There are 6 objects with a near-infrared 
excess but no submillimeter detection which are addressed in the Appendix.  The 
4 objects with no infrared excess but a submillimeter detection imply that the 
timescales for dissipation of the inner and outer disk are similar.  
\label{ccd}}
\end{figure}

\clearpage

\begin{figure}
\epsscale{0.9}
\plotone{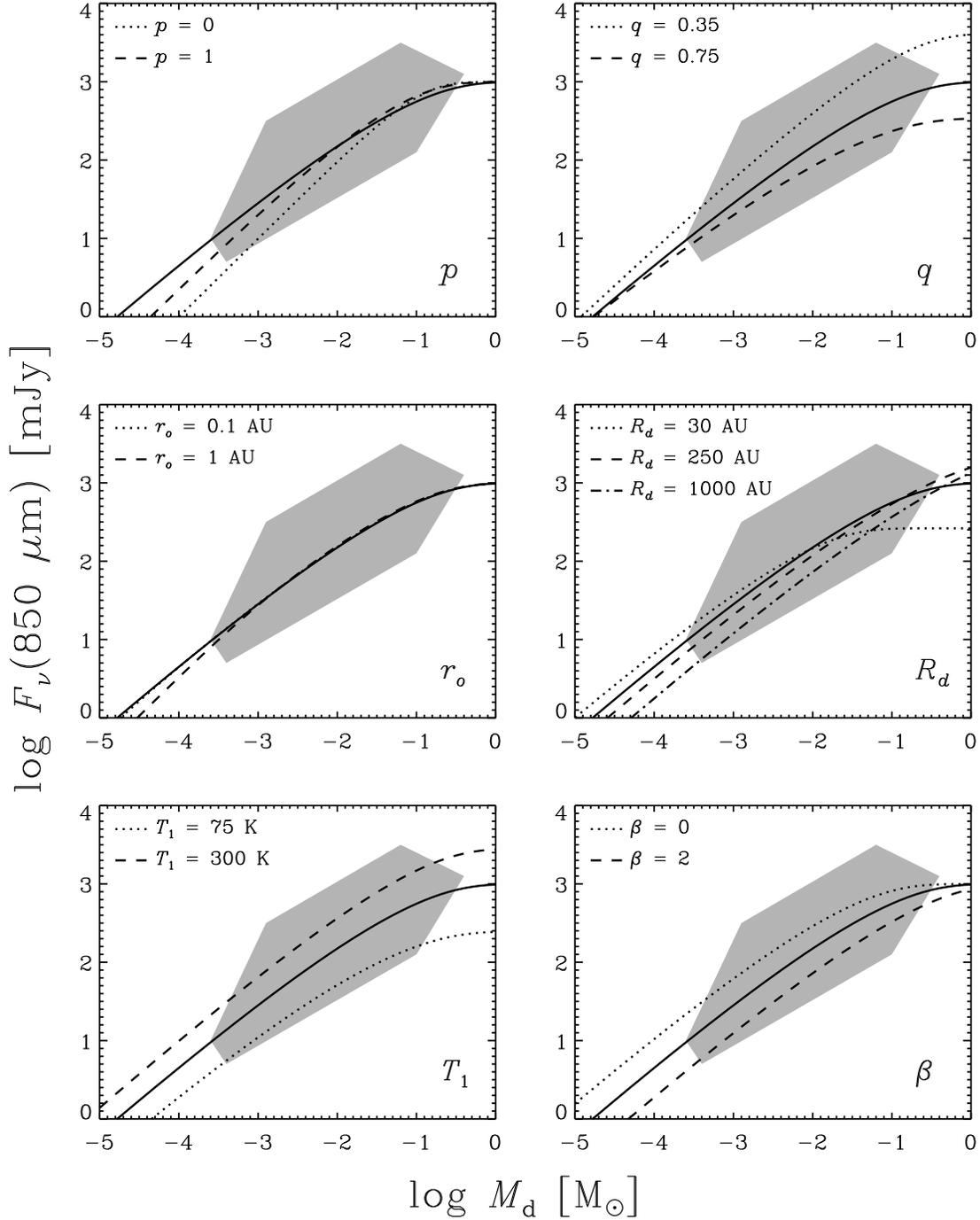}
\figcaption{The effects of reasonable variations of the SED model parameters on 
the relationship between the 850\,$\mu$m flux density and the disk mass.  The 
solid curve in each panel denotes the relationship for a fiducial parameter 
set: $i = 0$\degr, $p = 1.5$, $q = 0.6$, $r_{\circ} = 0.01$\,AU, $R_d = 
100$\,AU, $T_1 = 150$\,K, and $\beta = 1$.  The parameter being varied from 
this fiducial set is indicated in the lower right corner, and the different 
curves are marked in each panel.  The shaded area is representative of the data 
values (see Figure \ref{Md_F850}).  The temperature profile plays the dominant 
role in determining the relationship between $F_{\nu}$ and $M_d$. 
\label{FMparams}} 
\end{figure}

\clearpage

\begin{figure}
\epsscale{0.9}
\plotone{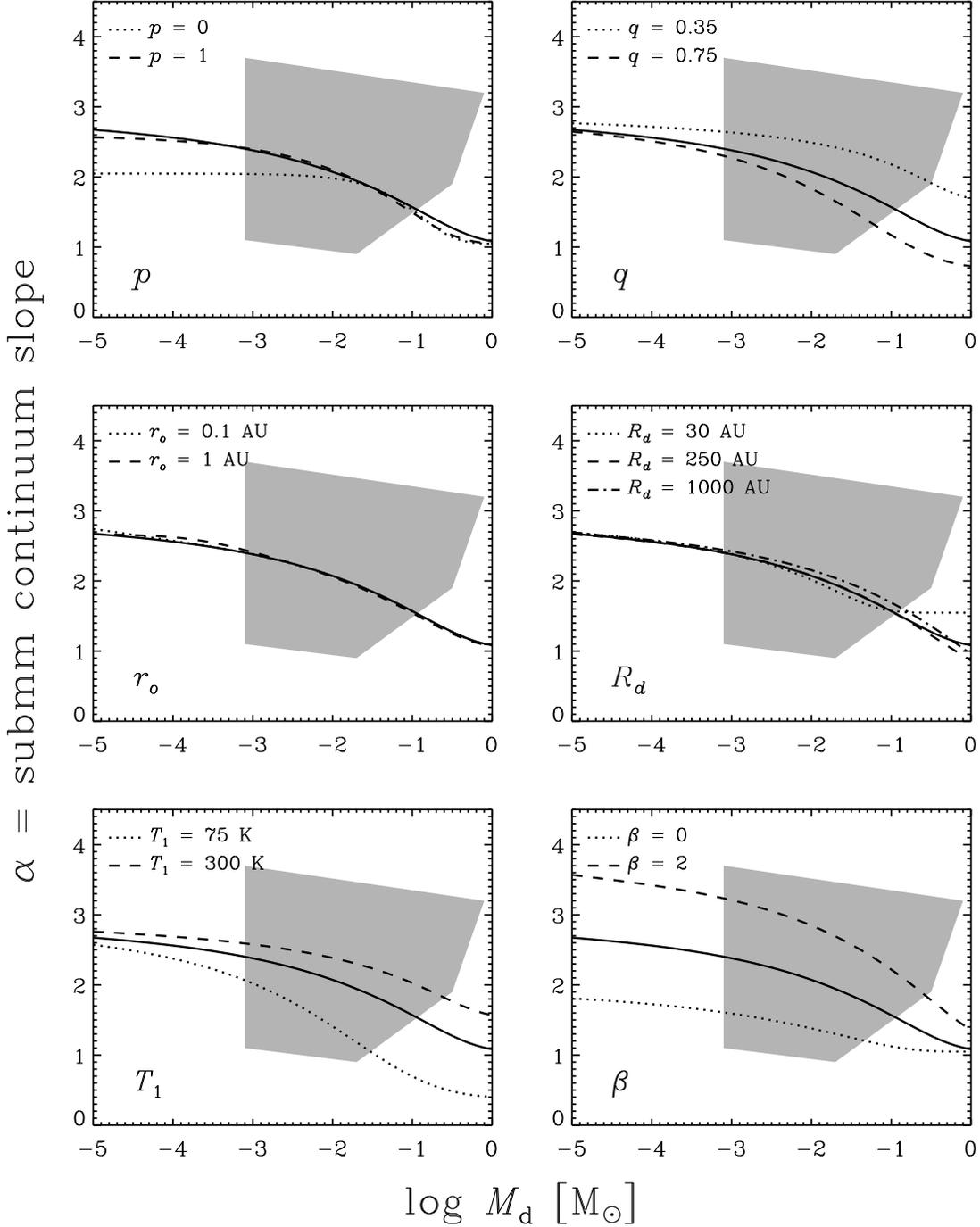}
\figcaption{The effects of reasonable variations of the SED model parameters on 
the relationship between the submillimeter continuum slope and the disk mass 
for a fixed opacity index $\beta = 1$.  The solid curve in each panel denotes 
the relationship for a fiducial parameter set: $i = 0$\degr, $p = 1.5$, $q = 
0.6$, $r_{\circ} = 0.01$\,AU, $R_d = 100$\,AU, $T_1 = 150$\,K, and $\beta = 
1$.  The parameter being varied from this fiducial set is indicated in the 
lower left corner, and the different curves are marked in each panel.  The 
shaded area is representative of the data values.  The lower right panel shows 
that $\beta$ is the dominant factor in setting the $\alpha - M_d$ relationship, 
but the temperature profile also has a significant effect.  \label{aMparams}}
\end{figure}

\clearpage

\begin{figure}
\epsscale{0.55}
\plotone{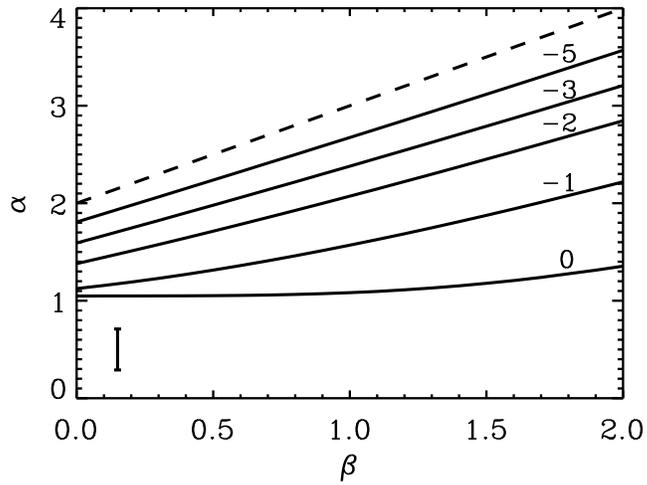}
\figcaption{Submillimeter continuum slopes ($\alpha$: defined by $F_{\nu} 
\propto \nu^{\alpha}$) between 350\,$\mu$m and 1.3\,mm from the fiducial disk 
model as a function of the power law index of the opacity ($\beta$) for various 
input disk masses.  Each curve is labeled with the logarithm of the disk mass.  
The dashed line marks the nominal relationship $\beta = \alpha - 2$ for 
optically thin emission in the Rayleigh-Jeans limit.  The relationship between 
$\beta$ and $\alpha$ is linear for disk masses less than 
$\sim$0.1\,M$_{\odot}$, but decreased relative to the optically thin case due 
to the fraction of emission (particularly at the shortest wavelengths) which is 
optically thick.  The error bar in the lower left corner shows the systematic 
uncertainty in $\alpha$ from absolute flux calibration errors.  \label{ab}}
\end{figure}

\clearpage

\begin{figure}
\epsscale{0.5}
\plotone{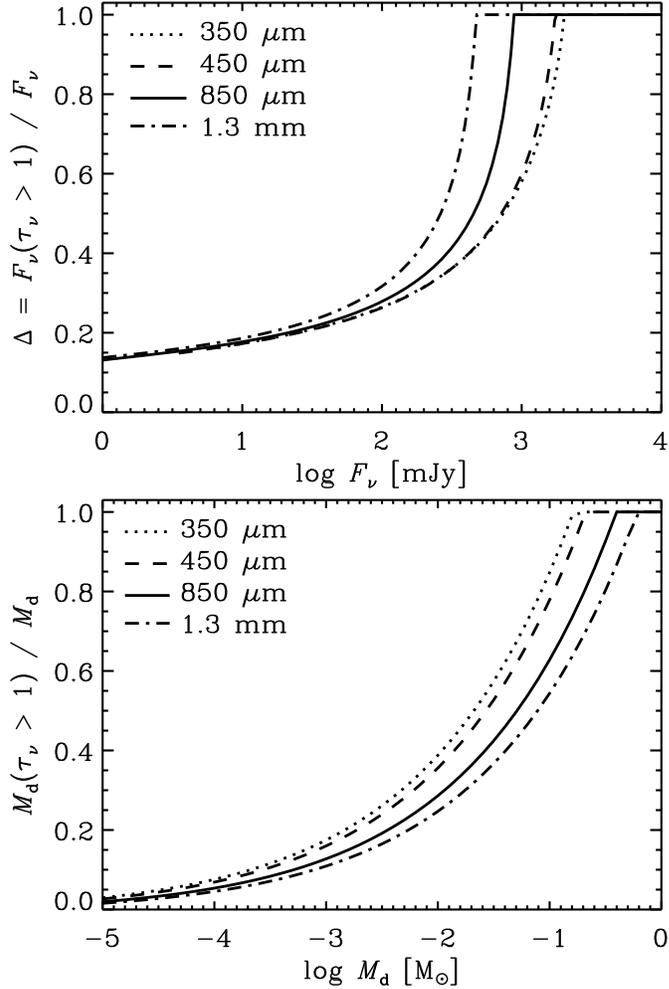}
\figcaption{(\emph{top}): The fraction of the submillimeter flux density 
which is from optically thick emission ($\Delta$) as a function of the total 
flux density for the fiducial disk model.  The various curves represent 
different wavelengths: 1.3\,mm (dash-dotted), 850\,$\mu$m (solid), 450\,$\mu$m 
(dashed), and 350\,$\mu$m (dotted).  (\emph{bottom}): The fraction of the disk 
mass which gives rise to optically thick submillimeter continuum emission at 
various wavelengths as a function of the total disk mass.  \label{tau}}
\end{figure} 

\clearpage

\begin{deluxetable}{lccrrrrlrc}
\tablecolumns{10}
\tabletypesize{\scriptsize}
\tablewidth{0pc}
\tablecaption{Submillimeter Properties of Taurus-Auriga Disks\tablenotemark{a} \label{results_table}}
\tablehead{
\colhead{} & \colhead{} & \colhead{} & \multicolumn{4}{c}{$F_{\nu}$ [mJy]} & \colhead{} & \colhead{} &\colhead{} \\ \cline{4-7} \colhead{Object} & \colhead{SED} & \colhead{SpT} & \colhead{$\lambda = 350$\,$\mu$m} & \colhead{$\lambda = 450$\,$\mu$m} & \colhead{$\lambda = 850$\,$\mu$m} & \colhead{$\lambda = 1.3$\,mm} & \colhead{$M_d$ [M$_{\odot}$]} & \colhead{$\alpha$} & \colhead{notes\tablenotemark{b}} \\ \colhead{(1)} & \colhead{(2)} & \colhead{(3)} & \colhead{(4)} & \colhead{(5)} & \colhead{(6)} & \colhead{(7)} & \colhead{(8)} & \colhead{(9)} & \colhead{(10)}} 
\startdata
04016+2610         & I   & \nodata & $12477 \pm 193$  & \nodata          & \nodata        & $130 \pm 5$   & 0.02      & $3.48 \pm 0.24$ & 5,8     \\
04113+2758         & II  & M2      & $7621 \pm 219$   & \nodata          & \nodata        & $410 \pm 40$  & 0.09      & $2.23 \pm 0.24$ & 5       \\
04154+2823         & FS  & \nodata & $440 \pm 83$     & $495 \pm 185$    & $140 \pm 6$    & \nodata       & 0.006     & $1.39 \pm 0.27$ & \nodata \\
04166+2706         & I   & \nodata & $6937 \pm 93$    & \nodata          & \nodata        & $180 \pm 8$   & 0.03      & $2.78 \pm 0.24$ & 5,8     \\
04169+2702         & I   & \nodata & $7344 \pm 152$   & \nodata          & \nodata        & $190 \pm 9$   & 0.03      & $2.79 \pm 0.24$ & 5,8     \\
04239+2436         & I   & \nodata & $1144 \pm 20$    & $<660$           & $207 \pm 9$    & $80 \pm 10$   & 0.009     & $2.01 \pm 0.24$ & 5,8     \\
04248+2612         & I   & M2      & $1178 \pm 30$    & \nodata          & \nodata        & $60 \pm 7$    & 0.005     & $2.27 \pm 0.21$ & 5,8     \\
04260+2642         & I   & K6      & $546 \pm 21$     & \nodata          & \nodata        & $105 \pm 10$  & 0.01      & $1.26 \pm 0.24$ & 5       \\
04278+2253         & II  & F1      & \nodata          & $<687$           & $36 \pm 7$     & \nodata       & 0.002     & $<4.64$         & \nodata \\
04295+2251         & FS  & \nodata & $1338 \pm 25$    & \nodata          & \nodata        & $115 \pm 10$  & 0.01      & $1.87 \pm 0.24$ & 5,8     \\
04301+2608         & II  & \nodata & \nodata          & $<351$           & $18 \pm 6$     & \nodata       & 0.0009    & $<4.67$         & \nodata \\
04302+2247         & I   & \nodata & $2869 \pm 21$    & \nodata          & \nodata        & $180 \pm 10$  & 0.03      & $2.11 \pm 0.24$ & 5,8     \\
04325+2402         & I   & \nodata & \nodata          & $606 \pm 185$    & $186 \pm 11$   & $110 \pm 7$   & 0.008     & $1.56 \pm 0.34$ & 5,8     \\
04361+2547         & I   & \nodata & \nodata          & $1302 \pm 168$   & $275 \pm 8$    & $110 \pm 8$   & 0.01      & $2.33 \pm 0.30$ & 5,8     \\
04365+2535         & I   & \nodata & \nodata          & $2928 \pm 230$   & $622 \pm 13$   & $230 \pm 10$  & 0.03      & $2.40 \pm 0.30$ & 5,8     \\
04368+2557         & I   & \nodata & \nodata          & $2849 \pm 222$   & $895 \pm 11$   & \nodata       & 0.04      & $1.82 \pm 0.42$ & 8       \\
04381+2540         & I   & \nodata & \nodata          & $1152 \pm 279$   & $208 \pm 11$   & $70 \pm 9$    & 0.009     & $2.64 \pm 0.30$ & 5,8     \\
AA Tau             & II  & K7      & $825 \pm 50$     & $415 \pm 84$     & $144 \pm 5$    & $88 \pm 9$    & 0.01      & $1.56 \pm 0.20$ & 3,6     \\
AB Aur             & II  & A0      & $8930 \pm 1410$  & $3820 \pm 570$   & $359 \pm 67$   & $103 \pm 18$  & 0.004     & $3.47 \pm 0.20$ & 2       \\
Anon 1             & III & M0      & \nodata          & $<79$            & $<8$           & $<14$         & $<0.0004$ & \nodata         & 4       \\
BP Tau             & II  & K7      & \nodata          & $<456$           & $130 \pm 7$    & $47 \pm 0.7$  & 0.02      & $2.39 \pm 0.53$ & 2       \\
CIDA-2             & III & M5      & \nodata          & $<165$           & $<14$          & \nodata       & $<0.0007$ & \nodata         & \nodata \\
CIDA-3             & II  & M2      & \nodata          & $<94$            & $<9$           & \nodata       & $<0.0004$ & \nodata         & \nodata \\
CIDA-7             & II  & M3      & \nodata          & $990 \pm 330$    & $38 \pm 8$     & \nodata       & 0.002     & $5.13 \pm 0.62$ & 7       \\
CIDA-8             & II  & M4      & \nodata          & $<80$            & $27 \pm 3$     & \nodata       & 0.001     & $<1.73$         & \nodata \\
CIDA-9             & II  & M0      & \nodata          & $843 \pm 259$    & $71 \pm 7$     & \nodata       & 0.003     & $3.89 \pm 0.51$ & \nodata \\
CIDA-10            & III & M4      & \nodata          & $<94$            & $<11$          & \nodata       & $<0.0005$ & \nodata         & \nodata \\
CIDA-11            & II  & M3      & \nodata          & $<87$            & $<8$           & \nodata       & $<0.0004$ & \nodata         & \nodata \\
CIDA-12            & II  & M4      & \nodata          & $<88$            & $<7$           & \nodata       & $<0.0004$ & \nodata         & \nodata \\
CI Tau             & II  & K7      & $1725 \pm 55$    & $846 \pm 89$     & $324 \pm 6$    & $190 \pm 17$  & $0.03$    & $1.79 \pm 0.21$ & 3,6     \\
CoKu Tau/1         & II  & M0      & \nodata          & $<522$           & $35 \pm 7$     & $<12$         & $0.002$   & $<4.25$         & 4       \\
CoKu Tau/3         & II  & M1      & \nodata          & $<104$           & $<8$           & $<16$         & $<0.0004$ & \nodata         & 4       \\
CoKu Tau/4         & II  & M2      & \nodata          & $<131$           & $9.0 \pm 2.9$  & $<15$         & 0.0005    & $<4.21$         & 4       \\
CW Tau             & II  & K2      & $1230 \pm 102$   & $<312$           & $66 \pm 6$     & $96 \pm 8$    & 0.002     & $2.09 \pm 0.24$ & 3,6     \\
CX Tau             & II  & M3      & \nodata          & $<570$           & $25 \pm 6$     & $<40$         & 0.001     & $<4.92$         & 3       \\
CY Tau             & II  & M1      & $<1839$          & $<210$           & $140 \pm 5$    & $133 \pm 11$  & 0.006     & $0.75 \pm 0.56$ & 3,6,7   \\
CZ Tau             & II  & M2      & \nodata          & $<262$           & $<9$           & $<30$         & $<0.0004$ & \nodata         & 3       \\
DD Tau             & II  & M1      & \nodata          & \nodata          & $<42$          & $17 \pm 4$    & 0.0007    & \nodata         & 4       \\
DE Tau             & II  & M2      & \nodata          & $<291$           & $90 \pm 7$     & $36 \pm 5$    & 0.005     & $2.16 \pm 0.40$ & 4       \\
DF Tau             & II  & M1      & \nodata          & $<304$           & $8.8 \pm 1.9$  & $<25$         & 0.0004    & $<5.57$         & 1,3     \\
DG Tau             & FS  & K7      & $5173 \pm 94$    & $3950 \pm 350$   & $1100 \pm 100$ & $700 \pm 130$ & 0.02      & $2.02 \pm 0.12$ & 2,6     \\
DH Tau             & II  & M1      & $261 \pm 9$      & \nodata          & $57 \pm 9$     & $<57$         & 0.003     & $1.71 \pm 0.33$ & 3       \\
DK Tau             & II  & K7      & \nodata          & $<419$           & $80 \pm 10$    & $35 \pm 7$    & 0.005     & $1.95 \pm 0.55$ & 3       \\
DL Tau             & II  & K7      & $1390 \pm 180$   & $1280 \pm 170$   & $440 \pm 40$   & $230 \pm 14$  & 0.09      & $1.54 \pm 0.15$ & 2,3,6   \\
DM Tau             & II  & M1      & $1077 \pm 49$    & \nodata          & $237 \pm 12$   & $109 \pm 13$  & 0.02      & $1.74 \pm 0.24$ & 3       \\
DN Tau             & II  & M0      & $615 \pm 64$     & $<703$           & $201 \pm 7$    & $84 \pm 13$   & 0.03      & $1.53 \pm 0.23$ & 3,6     \\
DO Tau             & II  & M0      & \nodata          & $734 \pm 50$     & $258 \pm 42$   & $136 \pm 11$  & 0.007     & $1.66 \pm 0.19$ & 1,3,6   \\
DP Tau             & II  & M1      & \nodata          & $<279$           & $<10$          & $<27$         & $<0.0005$ & \nodata         & 3       \\
DQ Tau             & II  & M0      & $244 \pm 76$     & $<861$           & $208 \pm 8$    & $91 \pm 9$    & 0.02      & $1.21 \pm 0.27$ & 3,6     \\
DR Tau             & II  & \nodata & \nodata          & $2380 \pm 172$   & $533 \pm 7$    & $159 \pm 11$  & 0.02      & $2.07 \pm 0.19$ & 1,3,6   \\
DS Tau             & II  & K5      & \nodata          & $<342$           & $39 \pm 4$     & $25 \pm 6$    & 0.006     & $1.05 \pm 0.61$ & 1,4,7   \\
FF Tau             & III & K7      & \nodata          & $<43$            & $<4$           & $<27$         & $<0.0002$ & \nodata         & 1,3     \\
FM Tau             & II  & M0      & $<349$           & $<442$           & $32 \pm 8$     & $<36$         & 0.002     & $<4.13$         & 3       \\
FO Tau             & II  & M2      & \nodata          & $<199$           & $13 \pm 3$     & $<14$         & 0.0006    & $<4.29$         & 1,4     \\
FQ Tau             & II  & M2      & \nodata          & $574 \pm 170$    & $28 \pm 7$     & $<40$         & 0.001     & $4.75 \pm 0.61$ & 3,7     \\
FS Tau             & FS  & M1      & \nodata          & $<576$           & $49 \pm 6$     & $<35$         & 0.002     & $<3.87$         & 3       \\
FT Tau             & II  & \nodata & $1106 \pm 82$    & $437 \pm 56$     & $121 \pm 5$    & $130 \pm 14$  & 0.01      & $1.66 \pm 0.20$ & 3,6     \\
FV Tau             & II  & K5      & \nodata          & $402 \pm 90$     & $48 \pm 5$     & $15 \pm 4$    & 0.001     & $3.15 \pm 0.34$ & 1,4     \\
FV Tau/c           & II  & M4      & \nodata          & $<355$           & $<25$          & $<16$         & $<0.001$  & \nodata         & 4       \\
FW Tau             & III & M4      & \nodata          & $<35$            & $4.5 \pm 1.1$  & $<15$         & 0.0002    & $<3.23$         & 1,3     \\
FX Tau             & II  & M1      & \nodata          & $<169$           & $17 \pm 3$     & $<30$         & 0.0009    & $<3.62$         & 3       \\
FY Tau             & II  & K7      & \nodata          & $<297$           & $<27$          & $16 \pm 5$    & 0.0007    & \nodata         & 3       \\
FZ Tau             & II  & M0      & \nodata          & $<273$           & $29 \pm 7$     & $23 \pm 7$    & 0.002     & $0.55 \pm 0.89$ & 3,7     \\
GG Tau             & II  & K7      & $6528 \pm 153$   & $2726 \pm 250$   & $1255 \pm 57$  & $593 \pm 53$  & 0.2       & $1.91 \pm 0.12$ & 1,3,6   \\
GH Tau             & II  & M2      & \nodata          & $<309$           & $15 \pm 3$     & $<30$         & 0.0007    & $<4.76$         & 1,3     \\
GK Tau             & II  & K7      & \nodata          & \nodata          & $33 \pm 7$     & $<21$         & 0.002     & \nodata         & 3       \\
GM Aur             & II  & K3      & $3419 \pm 133$   & \nodata          & \nodata        & $253 \pm 12$  & 0.03      & $2.25 \pm 0.23$ & 3,6     \\
GN Tau             & II  & \nodata & \nodata          & $<187$           & $12 \pm 3$     & $<50$         & 0.0006    & $<4.32$         & 1,3     \\
GO Tau             & II  & M0      & $274 \pm 26$     & $594 \pm 185$    & $173 \pm 7$    & $83 \pm 12$   & 0.07      & $1.77 \pm 0.33$ & 3,6     \\
GV Tau             & I   & K3      & $1676 \pm 137$   & $1808 \pm 121$   & $282 \pm 5$    & $87 \pm 4$    & 0.003     & $2.51 \pm 0.20$ & 4,6     \\
Haro 6-13          & FS  & M0      & $2729 \pm 171$   & $1400 \pm 180$   & $395 \pm 56$   & $124 \pm 13$  & 0.01      & $2.15 \pm 0.18$ & 1,3,6   \\
Haro 6-28          & FS  & M5      & \nodata          & $<2636$          & $11 \pm 3$     & $<14$         & 0.0006    & $<8.55$         & 1,4     \\
Haro 6-37          & II  & K6      & \nodata          & $536 \pm 204$    & $245 \pm 7$    & $<88$         & 0.01      & $1.23 \pm 0.62$ & 4       \\
Haro 6-39          & II  & \nodata & \nodata          & $<903$           & $36 \pm 6$     & $24 \pm 6$    & 0.002     & $0.95 \pm 0.72$ & 3,7     \\
HBC 347            & III & K1      & \nodata          & $<138$           & $<9$           & \nodata       & $<0.0004$ & \nodata         & 4       \\
HBC 351	           & III & K5      & \nodata          & $<166$           & $<11$          & $<14$         & $<0.0005$ & \nodata         & 4       \\
HBC 352/353        & III & G0      & \nodata          & $<81$            & $<9$           & $<12$         & $<0.0005$ & \nodata         & 4       \\
HBC 354/355        & III & K3      & \nodata	      & $<71$            & $<7$           & $<16$         & $<0.0004$ & \nodata         & 4       \\
HBC 356/357        & III & K2      & \nodata          & $<69$            & $<9$           & $<16$         & $<0.0004$ & \nodata         & 4       \\
HBC 358/359        & III & M2      & \nodata          & $<72$            & $<9$           & $<13$         & $<0.0005$ & \nodata         & 4       \\
HBC 360/361        & III & M3      & \nodata          & $<243$           & $<14$          & $<26$         & $<0.0007$ & \nodata         & 1,4     \\
HBC 362            & III & M2      & \nodata          & $<108$           & $<8$           & $<14$         & $<0.0004$ & \nodata         & 4       \\
HBC 372            & III & K5      & \nodata          & $<173$           & $<8$           & $<14$         & $<0.0004$ & \nodata         & 4       \\
HBC 376            & III & K7      & \nodata          & $<49$            & $<6$           & $<14$         & $<0.0003$ & \nodata         & 1,4     \\
HBC 388            & III & K1      & \nodata          & $<54$            & $<6$           & $<16$         & $<0.0003$ & \nodata         & 1,4     \\
HBC 392            & III & K5      & \nodata          & $<141$           & $<6$           & $<18$         & $<0.0003$ & \nodata         & 1,4     \\
HBC 407            & III & G8      & \nodata          & $<146$           & $<9$           & $<64$         & $<0.0004$ & \nodata         & 4       \\
HBC 412            & III & M2      & \nodata          & $<200$           & $<9$           & $<16$         & $<0.0004$ & \nodata         & 4       \\
HBC 427            & III & K7      & \nodata          & $<1516$          & $<14$          & \nodata       & $<0.0007$ & \nodata         & 1       \\
HD 283572          & III & G5      & \nodata          & $<216$           & $<9$           & $<15$         & $<0.0004$ & \nodata         & 4       \\
HD 283759          & III & F3      & \nodata          & $<149$           & $<10$          & $<35$         & $<0.0005$ & \nodata         & 3       \\
HK Tau             & FS  & M1      & $680 \pm 114$    & \nodata          & \nodata        & $41 \pm 5$    & 0.004     & $2.14 \pm 0.21$ & 3,6     \\
HL Tau             & I   & K7      & $23888 \pm 149$  & $10400 \pm 1400$ & $2360 \pm 90$  & $880 \pm 19$  & 0.06      & $2.53 \pm 0.13$ & 2,3,6,8 \\
HN Tau             & II  & K5      & \nodata          & $<171$           & $29 \pm 3$     & $<15$         & 0.0008    & $<2.79$         & 1,4     \\
HO Tau             & II  & M1      & \nodata          & $<567$           & $44 \pm 6$     & $<30$         & 0.002     & $<4.02$         & 3       \\
HP Tau             & FS  & K3      & $386 \pm 63$     & \nodata          & \nodata        & $62 \pm 6$    & 0.001     & $1.39 \pm 0.24$ & 3       \\
HQ Tau             & III & \nodata & \nodata          & $<221$           & $11 \pm 3$     & $<45$         & 0.0005    & $<4.72$         & 1,3     \\
Hubble 4           & III & K7      & \nodata          & $<89$            & $<9$           & $<25$         & $<0.0004$ & \nodata         & 3       \\
HV Tau             & II  & M1      & \nodata          & $<519$           & $47 \pm 4$     & $40 \pm 6$    & 0.002     & $0.38 \pm 0.42$ & 1,4,7   \\
IC 2087/IR         & II  & \nodata & \nodata          & $1365 \pm 130$   & $501 \pm 7$    & \nodata       & 0.02      & $1.58 \pm 0.42$ & \nodata \\
IP Tau             & II  & M0      & \nodata          & $<516$           & $34 \pm 5$     & $16 \pm 5$    & 0.003     & $1.80 \pm 0.81$ & 4,7     \\
IQ Tau             & II  & M1      & \nodata          & $425 \pm 26$     & $178 \pm 3$    & $87 \pm 11$   & 0.02      & $1.53 \pm 0.30$ & 3,6     \\
IS Tau             & II  & K7      & \nodata          & $<252$           & $30 \pm 3$     & $<20$         & 0.001     & $<3.35$         & 1,3     \\
IT Tau             & II  & K2      & \nodata          & $<73$            & $22 \pm 3$     & $<33$         & 0.002     & $<1.86$         & 3       \\
IW Tau             & III & K7      & \nodata          & $<253$           & $<9$           & $<19$         & $<0.0004$ & \nodata         & 1,4     \\
J1-4423            & III & M5      & \nodata          & $<52$            & $<8$           & $<11$         & $<0.0004$ & \nodata         & 4       \\
J1-4872            & III & K7      & \nodata          & $<63$            & $<8$           & $<14$         & $<0.0004$ & \nodata         & 4       \\
J1-507             & III & M4      & \nodata          & $<52$            & $<6$           & $<14$         & $<0.0003$ & \nodata         & 4       \\
JH 56              & III & M1      & \nodata          & $<74$            & $<8$           & $<19$         & $<0.0004$ & \nodata         & 4       \\
JH 108             & III & M1      & \nodata          & $<66$            & $<7$           & $<18$         & $<0.0004$ & \nodata         & 4       \\
JH 112             & II  & K6      & \nodata          & \nodata          & $30 \pm 10$    & $<18$         & 0.001     & $<5.51$         & 4       \\
JH 223             & II  & M2      & \nodata          & $<62$            & $<7$           & $<19$         & $<0.0003$ & \nodata         & 4       \\
L1551-51           & III & K7      & \nodata          & $<201$           & $<13$          & \nodata       & $<0.0006$ & \nodata         & \nodata \\
L1551-55           & III & K7      & \nodata          & $<60$            & $<5$           & $<23$         & $<0.0003$ & \nodata         & 1,4     \\
L1551 IRS5         & I   & \nodata & $100423 \pm 812$ & \nodata          & \nodata        & $1276 \pm 5$  & 0.5       & $2.95 \pm 0.17$ & 4,8     \\
L1551 NE           & I   & \nodata & $22826 \pm 715$  & \nodata          & \nodata        & $850 \pm 10$  & 0.3       & $2.51 \pm 0.24$ & 5       \\
LkCa 1             & III & M4      & \nodata          & $<89$            & $<8$           & $<14$         & $<0.0004$ & \nodata         & 4       \\
LkCa 3             & III & M1      & \nodata          & $<471$           & $<9$           & $<14$         & $<0.0004$ & \nodata         & 1,4     \\
LkCa 4             & III & K7      & \nodata          & $<37$            & $<4$           & $<14$         & $<0.0002$ & \nodata         & 1,4     \\
LkCa 5             & III & M2      & \nodata          & $<28$            & $<4$           & $<14$         & $<0.0002$ & \nodata         & 1,4     \\
LkCa 7             & III & K7      & \nodata          & $<107$           & $<9$           & \nodata       & $<0.0004$ & \nodata         & \nodata \\
LkCa 14            & III & M0      & \nodata          & $<103$           & $<9$           & $<19$         & $<0.0004$ & \nodata         & 4       \\
LkCa 15            & II  & K5      & $1235 \pm 80$    & \nodata          & $428 \pm 11$   & $167 \pm 6$   & 0.05      & $1.49 \pm 0.24$ & 4       \\
LkCa 19            & III & K0      & \nodata          & $<90$            & $<10$          & \nodata       & 0.0005    & \nodata         & \nodata \\
LkCa 21            & III & M3      & \nodata          & $<145$           & $<10$          & $<12$         & $<0.0005$ & \nodata         & 4       \\
LkH$\alpha$ 332/G1 & III & M1      & \nodata          & $<663$           & $12 \pm 3$     & $<14$         & 0.0006    & $<6.38$         & 1,4     \\
LkH$\alpha$ 332/G2 & III & K7      & \nodata          & $<1083$          & $<9$           & $<15$         & $<0.0005$ & \nodata         & 1,4     \\
RW Aur             & II  & K3      & $305 \pm 32$     & $167 \pm 34$     & $79 \pm 4$     & $42 \pm 5$    & 0.004     & $1.43 \pm 0.21$ & 4       \\
RY Tau             & II  & K1      & $2439 \pm 330$   & $1920 \pm 160$   & $560 \pm 30$   & $229 \pm 17$  & 0.02      & $1.79 \pm 0.17$ & 2,3,6   \\
SAO 76411          & III & G1      & \nodata          & $<102$           & $<9$           & $<14$         & $<0.0005$ & \nodata         & 4       \\
SAO 76428          & III & F8      & \nodata          & $<123$           & $<12$          & $<14$         & $<0.0006$ & \nodata         & 4       \\
St 34              & II  & M3      & \nodata          & $<243$           & $<11$          & $<15$         & $<0.0005$ & \nodata         & 4       \\
SU Aur             & II  & G2      & \nodata          & $251 \pm 40$     & $74 \pm 3$     & $<30$         & 0.0009    & $1.73 \pm 0.21$ & 3       \\
T Tau              & II  & K0      & $8149 \pm 253$   & $1655 \pm 218$   & $628 \pm 17$   & $280 \pm 9$   & 0.008     & $2.13 \pm 0.30$ & 1,3,6,8 \\
UX Tau             & II  & K2      & \nodata          & $523 \pm 37$     & $173 \pm 3$    & $63 \pm 10$   & 0.005     & $2.00 \pm 0.30$ & 4       \\
UY Aur             & II  & K7      & $542 \pm 77$     & $<523$           & $102 \pm 6$    & $29 \pm 6$    & 0.002     & $2.19 \pm 0.24$ & 4       \\
UZ Tau             & II  & M1      & $1823 \pm 142$   & $1811 \pm 129$   & $560 \pm 7$    & $172 \pm 15$  & 0.02      & $1.92 \pm 0.20$ & 1,3,6   \\
V410 Tau           & III & K3      & \nodata          & $<206$           & $7.2 \pm 1.8$  & $<30$         & $<0.0004$ & $<5.27$         & 1,3,7   \\
V710 Tau           & II  & M1      & \nodata          & $<495$           & $152 \pm 6$    & $60 \pm 7$    & 0.007     & $2.19 \pm 0.36$ & 4,6     \\
V773 Tau           & II  & K3      & \nodata          & $<386$           & $9.2 \pm 2.9$  & $24 \pm 4$    & 0.0005    & $<5.88$         & 1,4,7   \\
V807 Tau           & III & K7      & \nodata          & $<202$           & $20 \pm 3$     & $<18$         & 0.001     & $<3.64$         & 1,4     \\
V819 Tau           & III & K7      & \nodata          & $<317$           & $<9$           & $<9$          & $<0.0004$ & \nodata         & 1,4     \\
V826 Tau           & III & M0      & \nodata          & $<234$           & $<7$           & $<15$         & $<0.0004$ & \nodata         & 1,4     \\
V827 Tau           & III & K7      & \nodata          & $<147$           & $<6$           & $<19$         & $<0.0003$ & \nodata         & 1,4     \\
V830 Tau           & III & M0      & \nodata          & $<57$            & $<6$           & $<9$          & $<0.0003$ & \nodata         & 1,4     \\
V836 Tau           & II  & K7      & $344 \pm 30$     & $231 \pm 43$     & $74 \pm 3$     & $37 \pm 6$    & 0.01      & $1.70 \pm 0.24$ & 4       \\
V892 Tau           & II  & A0      & $4100 \pm 560$   & $2570 \pm 350$   & $638 \pm 54$   & $234 \pm 19$  & 0.009     & $2.20 \pm 0.19$ & 2,6     \\
V927 Tau           & III & M5      & \nodata          & $<1030$          & $<10$          & $<20$         & $<0.0005$ & \nodata         & 1,3     \\
V928 Tau           & III & M1      & \nodata          & $<258$           & $<8$           & $<11$         & $<0.0004$ & \nodata         & 1,4     \\
V955 Tau           & II  & M0      & \nodata          & $<390$           & $14 \pm 2$     & $<19$         & 0.0005    & $<5.27$         & 1,4     \\
VY Tau             & II  & M0      & \nodata          & $<225$           & $<10$          & $<17$         & $<0.0005$ & \nodata         & 1,4     \\
Wa Tau/1           & III & K0      & \nodata          & $<55$            & $<6$           & $<19$         & $<0.0003$ & \nodata         & 1,4     \\
ZZ Tau             & III & M3      & \nodata          & $<251$           & $<8$           & $<15$         & $<0.0004$ & \nodata         & 1,3     \\
\enddata
\tablenotetext{a}{The columns are as follows: (1) $-$ object name; (2) $-$ SED 
classification type (FS = Flat Spectrum); (3) $-$ spectral type from the 
literature (see \S 3.4); (4) $-$ 350\,$\mu$m flux density; (5) $-$ 450\,$\mu$m 
flux density; (6) $-$ 850\,$\mu$m flux density; (7) $-$ 1.3\,mm flux density 
values taken from the literature; (8) $-$ logarithm of the disk mass (see \S 
3.2); (9) $-$ submillimeter continuum slope (see \S 3.3); (10) $-$ notes on 
individual sources.  All flux densities are measured in units of mJy.  Upper 
limits are taken at the 3-$\sigma$ confidence level.  Quoted errors are the 
1-$\sigma$ rms noise levels and do not include systematic errors in the 
absolute flux calibration ($\sim$25\% at 350 and 450\,$\mu$m, $\sim$10\% at 
850\,$\mu$m, and $\sim$20\% at 1.3\,mm).}
\tablenotetext{b}{The numbers in the notes column (10) refer to the following 
information: 1 $-$ 450 and 850\,$\mu$m data are from the JCMT SCUBA archive.  
The original data were taken at various times between 1997 and 2002 and were 
reduced in the same way described in \S 2 with slight modifications for the 
different filter set before 1999 November.  2 $-$ Flux densities for V892 Tau 
(except 1.3\,mm) and AB Aur are from \citet{mannings94}.  Flux densities for HL 
Tau are taken from the long-term, repeated measurements at SCUBA \citep[450 and 
850\,$\mu$m: see \url{http://www.jach.hawaii.edu/JCMT/continuum/calibration/sens/secondary$\_$2004.html} and][]{jenness02}.  Flux densities for RY Tau (except 
1.3\,mm), DG Tau (except 350\,$\mu$m), and DL Tau (except 1.3\,mm) are taken 
from \citet{mannings94b}.  The 1.3\,mm flux density for BP Tau was taken from 
\citet{dutrey03}.  3 $-$ 1.3\,mm flux densities from \citet{bscg90}.  4 $-$ 
1.3\,mm flux densities from \citet{osterloh95}.  5 $-$ 1.3\,mm flux densities 
from \citet{motte01}.  6 $-$ Additional submillimeter flux densities 
\citep{adams90,beckwith91,mannings94} were used in determining $\alpha$.  7 $-$ 
These sources have anomalous continuum slopes, and so are excluded from the 
analysis in \S 3.3.  Two of these objects with 850\,$\mu$m detections are known 
to be variable centimeter radio sources (V773 Tau and V410 Tau).  Because V410 
Tau has no signatures of disk emission throughout its SED, we assume that its 
submillimeter flux is not from a dust disk, and instead use a 3-$\sigma$ upper 
limit of 5\,mJy at 850\,$\mu$m in all the analysis.  V773 Tau has a slight 
infrared excess, and so we consider the submillimeter emission to be from the 
disk.  The anomalous slopes for the other sources could be due to contamination 
from non-disk emission at the longest wavelengths, or errors in the absolute 
calibrations at different wavelengths.  See the Appendix regarding CY Tau.  8 
$-$ References for submillimeter maps of some Class I and FS YSOs in the 
literature.  See \citet{young03} for maps of 04016+2610, 04166+2706 (also 
\citet{shirley00}), 04169+2702, 04239+2436 (also \citet{chini01}), 04248+2612, 
04295+2251, 04302+2247, 04361+2547, and 04381+2540 (also 
\citet{hogerheijde00}).  See \citet{hogerheijde00} for maps of 04325+2402 and 
04368+2557 (also \citet{chini01}).  \citet{chandler00} provide maps of HL Tau 
and \citet{sandell01} provide maps of L1551 IRS 5.  Maps of the extended 
submillimeter emission around T Tau are provided by \citet{weintraub99}.}
\end{deluxetable}

\clearpage

\begin{deluxetable}{lccrccccc}
\tablecolumns{9}
\tabletypesize{\scriptsize}
\tablewidth{0pc}
\tablecaption{Results of SED Fits\tablenotemark{a} \label{fit_results}}
\tablehead{
\colhead{Object} & \colhead{$T_1$ [K]} & \colhead{$q$} & \colhead{$M_d$ [M$_{\odot}$]} & \colhead{$\Delta$} & \colhead{$\tilde{\chi}_{\nu}^2$} & \colhead{$\nu$} & \colhead{$T_c$ [K]} & \colhead{notes} \\ \colhead{(1)} & \colhead{(2)} & \colhead{(3)} & \colhead{(4)} & \colhead{(5)} & \colhead{(6)} & \colhead{(7)} & \colhead{(8)} & \colhead{(9)}}
\startdata
AA Tau      & 129 & 0.56 & $1.3 \pm 0.2 \times 10^{-2}$ & 0.31 & 4.7 & 8  & 15  & 1,2,3   \\
AB Aur      & 367 & 0.45 & $4.4 \pm 0.6 \times 10^{-3}$ & 0.17 & 2.6 & 6  & 65  & 1,2     \\
BP Tau      & 117 & 0.64 & $1.8 \pm 0.4 \times 10^{-2}$ & 0.42 & 2.0 & 4  & 11  & 1,2     \\
CI Tau      & 152 & 0.56 & $2.8 \pm 0.6 \times 10^{-2}$ & 0.36 & 1.7 & 9  & 15  & 1,2,3   \\
CW Tau      & 204 & 0.62 & $2.4 \pm 0.4 \times 10^{-3}$ & 0.24 & 5.7 & 6  & 27  & 2,3     \\
DE Tau      & 130 & 0.55 & $5.2 \pm 0.8 \times 10^{-3}$ & 0.25 & 0.6 & 3  & 19  & 1,2     \\
DF Tau      & 126 & 0.74 & $4 \pm 1 \times 10^{-4}$     & 0.28 & 3.6 & 3  & 23  & 1,2     \\
DG Tau      & 288 & 0.51 & $2.4 \pm 0.3 \times 10^{-2}$ & 0.29 & 2.7 & 13 & 40  & 1,2,3,4 \\
DH Tau      & 109 & 0.55 & $3.3 \pm 0.7 \times 10^{-3}$ & 0.24 & 2.4 & 4  & 19  & 1,2     \\
DK Tau      & 175 & 0.70 & $5 \pm 1 \times 10^{-3}$     & 0.34 & 0.6 & 4  & 18  & 1,2     \\
DL Tau      & 149 & 0.62 & $9 \pm 2 \times 10^{-2}$     & 0.58 & 0.6 & 12 & 10  & 2,3,4   \\
DM Tau      & 111 & 0.51 & $2.4 \pm 0.4 \times 10^{-2}$ & 0.34 & 2.0 & 6  & 14  & 2,3     \\
DN Tau      & 117 & 0.60 & $2.9 \pm 0.6 \times 10^{-2}$ & 0.43 & 1.5 & 5  & 11  & 2,3     \\
DO Tau      & 193 & 0.52 & $7 \pm 1 \times 10^{-3}$     & 0.23 & 2.3 & 12 & 33  & 1,2,3,4 \\
DQ Tau      & 143 & 0.60 & $1.9 \pm 0.5 \times 10^{-2}$ & 0.36 & 0.4 & 4  & 14  & 2,3     \\
DR Tau      & 216 & 0.58 & $1.9 \pm 0.3 \times 10^{-2}$ & 0.32 & 1.5 & 10 & 27  & 1,2,3,4 \\
DS Tau      & 97  & 0.67 & $6 \pm 1 \times 10^{-3}$     & 0.39 & 1.0 & 4  & 11  & 1,2     \\
FT Tau      & 121 & 0.58 & $1.4 \pm 0.2 \times 10^{-2}$ & 0.34 & 5.0 & 8  & 13  & 2,3     \\
FV Tau      & 190 & 0.53 & $1.1 \pm 0.2 \times 10^{-3}$ & 0.17 & 1.4 & 4  & 38  & 2       \\
FX Tau      & 118 & 0.63 & $9 \pm 2 \times 10^{-4}$     & 0.24 & 0.6 & 2  & 21  & 1,2     \\
FZ Tau      & 148 & 0.73 & $2.0 \pm 0.6 \times 10^{-3}$ & 0.33 & 1.9 & 2  & 17  & 2       \\
GG Tau      & 172 & 0.56 & $2.3 \pm 0.8 \times 10^{-1}$ & 0.76 & 1.0 & 8  & 10  & 2,3     \\
GM Aur      & 136 & 0.44 & $2.5 \pm 0.5 \times 10^{-2}$ & 0.28 & 2.1 & 6  & 30  & 2,3     \\
GO Tau      & 90  & 0.62 & $7 \pm 2 \times 10^{-2}$     & 0.62 & 1.2 & 7  & 7   & 2,3     \\
GV Tau      & 339 & 0.46 & $2.8 \pm 0.3 \times 10^{-3}$ & 0.16 & 1.5 & 7  & 78  & 2,3     \\
Haro 6-13   & 181 & 0.47 & $1.1 \pm 0.1 \times 10^{-2}$ & 0.23 & 0.9 & 11 & 33  & 2,3,4   \\
HK Tau      & 140 & 0.44 & $4.5 \pm 0.5 \times 10^{-3}$ & 0.18 & 2.8 & 9  & 28  & 1,2,3,5 \\
HL Tau      & 277 & 0.42 & $6.5 \pm 0.8 \times 10^{-2}$ & 0.36 & 0.7 & 12 & 33  & 1,2,3,5 \\
HN Tau      & 164 & 0.56 & $8 \pm 2 \times 10^{-4}$     & 0.18 & 1.9 & 2  & 33  & 1,2     \\
HP Tau      & 201 & 0.47 & $1.0 \pm 0.3 \times 10^{-3}$ & 0.14 & 5.9 & 4  & 160 & 1,2     \\
IP Tau      & 107 & 0.63 & $2.8 \pm 0.6 \times 10^{-3}$ & 0.29 & 1.1 & 3  & 15  & 2       \\
IQ Tau      & 121 & 0.60 & $2.2 \pm 0.3 \times 10^{-2}$ & 0.40 & 2.4 & 5  & 12  & 1,2,3   \\
IT Tau      & 104 & 0.61 & $1.5 \pm 0.3 \times 10^{-3}$ & 0.25 & 5.1 & 2  & 17  & 1,2     \\
LkCa 15     & 117 & 0.52 & $4.8 \pm 0.9 \times 10^{-2}$ & 0.41 & 2.2 & 5  & 13  & 1,2     \\
RY Tau      & 342 & 0.66 & $1.8 \pm 0.3 \times 10^{-2}$ & 0.34 & 1.4 & 13 & 52  & 1,2,3,4 \\
SU Aur      & 264 & 0.48 & $9 \pm 3 \times 10^{-4}$     & 0.14 & 3.2 & 4  & 66  & 1,2     \\
T Tau       & 338 & 0.45 & $8.2 \pm 0.9 \times 10^{-3}$ & 0.19 & 1.9 & 12 & 62  & 1,2,3,5 \\
UX Tau      & 132 & 0.41 & $5.1 \pm 0.7 \times 10^{-3}$ & 0.17 & 3.3 & 5  & 31  & 1,2     \\
UY Aur      & 226 & 0.53 & $1.8 \pm 0.3 \times 10^{-3}$ & 0.18 & 1.8 & 5  & 47  & 1,2     \\
UZ Tau      & 167 & 0.61 & $6 \pm 1 \times 10^{-2}$     & 0.49 & 2.1 & 9  & 13  & 1,2,3   \\
V710 Tau    & 112 & 0.58 & $1.9 \pm 0.4 \times 10^{-2}$ & 0.37 & 0.6 & 5  & 12  & 2,3     \\
V836 Tau    & 97  & 0.58 & $1.0 \pm 0.3 \times 10^{-2}$ & 0.34 & 0.6 & 5  & 12  & 2       \\
V892 Tau    & 461 & 0.58 & $9 \pm 1 \times 10^{-3}$     & 0.25 & 1.3 & 8  & 58  & 2,3     \\
V955 Tau    & 139 & 0.59 & $5 \pm 1 \times 10^{-4}$     & 0.19 & 6.2 & 2  & 27  & 2
\enddata
\tablenotetext{a}{The table columns are: (1) $-$ object name; (2) $-$ best-fit 
value of the temperature at 1\,AU in K (typical errors are $\pm$ a few K); (3) 
$-$ best-fit value of the radial power law index of the temperature profile 
(typical errors are $\pm$ 0.02); (4) $-$ best-fit value of the disk mass for 
$\beta = 1$; (5) $-$ fraction of the 850\,$\mu$m flux density from optically 
thick regions in the disk; (6) $-$ \emph{reduced} chi-squared statistic; (7) 
$-$ number of degrees of freedom in the fit (i.e., number of datapoints $-$ 
number of fitted parameters [= 3]); (8) $-$ inferred characteristic temperature 
in K from inversion of Equation 1; (9) $-$ notes on literature sources for the 
data as follows: 1 = 10\,$\mu$m photometry from \citet{kh95}, 2 = \emph{IRAS} 
photometry (12, 25, 60, 100\,$\mu$m) from \citet{weaver92}, 3 = submillimeter 
photometry from \citet{beckwith91}, 4 = submillimeter photometry from 
\citet{mannings94b}, 5 = submillimeter photometry from \citet{adams90}.  The 
largest values of $\tilde{\chi}_{\nu}^2$ usually are due to structure in the 
infrared SED, where the errors on flux densities are low, or slightly 
inconsistent absolute calibration in the submillimeter.}
\end{deluxetable}

\begin{deluxetable}{lccc|lccc}
\tablecolumns{8}
\tablewidth{0pc}
\tabletypesize{\scriptsize}
\tablecaption{Multiple Star Systems\tablenotemark{a} \label{binaries}}
\tablehead{
\colhead{Object} & \colhead{$\theta_p$ [\arcsec]} & \colhead{ref\tablenotemark{b}} & \colhead{$a_p$ [AU]} & \colhead{Object} & \colhead{$\theta_p$ [\arcsec]} & \colhead{ref\tablenotemark{b}} & \colhead{$a_p$ [AU]} \\ \colhead{(1)} & \colhead{(2)} & \colhead{(3)} & \colhead{(4)} & \colhead{(1)} & \colhead{(2)} & \colhead{(3)} & \colhead{(4)}} 
\startdata
04113+2758              & 4.0            & 18       & 560          & HK Tau                  & 2.4            & 7        & 340          \\
04325+2402              & 8.2            & 18       & 1140         & HN Tau                  & 3.1            & 8        & 430          \\
04248+2612              & 4.6            &  18      & 640          & HO Tau                  & 6.9            & 11       & 970          \\
CoKu Tau/3              & 2.1            & 7        & 290          & HP Tau                  & 0.02           & 10       & 3            \\
CZ Tau                  & 0.33           & 8        & 46           & HV Tau                  & 0.03, 4.0      & 6        & 4, 560       \\
DD Tau                  & 0.56           & 8        & 79           & IS Tau                  & 0.22           & 9        & 31           \\
DF Tau                  & 0.09           & 16       & 13           & IT Tau                  & 2.4            & 7        & 340          \\
DK Tau                  & 2.5            & 6        & 350          & IW Tau                  & 0.28           & 8        & 39           \\
DQ Tau                  & sb             & 14       & 0.05         & J1-4872                 & 3.3            & 16       & 460          \\
DS Tau                  & 7.1            & 11       & 990          & L1551 IRS 5             & 0.30           & 1        & 42           \\
FF Tau                  & 0.03           & 10       & 4            & LkCa 3                  & sb, 0.48       & 11, 16   & 0.03, 67     \\
FO Tau                  & 0.16           & 8        & 22           & LkCa 7                  & 1.0            & 8        & 140          \\
FQ Tau                  & 0.78           & 8        & 110          & LkH$\alpha$ 332/G1      & 0.22           & 8        & 31           \\
FS Tau                  & 0.25           & 8        & 35           & LkH$\alpha$ 332/G2      & 0.28           & 8        & 39           \\
FV Tau                  & 0.72           & 8        & 101          & RW Aur                  & 1.4            & 7        & 200          \\
FV Tau/c                & 0.74           & 6        & 104          & St 34                   & sb             & 20       & \nodata      \\
FW Tau                  & 0.16, 2.3      & 12, 16   & 22, 320      & T Tau\tablenotemark{c}  & 0.10, 0.70     & 17, 8    & 14, 98       \\
FX Tau                  & 0.90           & 9        & 130          & UX Tau\tablenotemark{c} & 0.14, 2.7, 5.9 & 15, 7, 7 & 20, 380, 830 \\
GG Tau\tablenotemark{d} & 0.26           & 8        & 36           & UY Aur                  & 0.88           & 9        & 120          \\
GH Tau                  & 0.33           & 9        & 46           & UZ Tau\tablenotemark{c} & sb, 0.35, 3.7  & 19, 9, 7 & 0.1, 49, 520 \\
GK Tau                  & 2.5            & 7        & 340          & V410 Tau                & 0.07, 0.29     & 16       & 10, 41       \\
GN Tau\tablenotemark{d} & 0.04           & 6        & 6            & V710 Tau                & 3.2            & 7        & 450          \\
GV Tau                  & 1.2            & 3        & 170          & V773 Tau                & 0.11           & 9        & 15           \\
Haro 6-28               & 0.66           & 8        & 92           & V807 Tau                & 0.02, 0.37     & 12, 9    & 3, 52        \\
Haro 6-37               & 2.7            & 8        & 370          & V826 Tau                & sb             & 0        & 0.05         \\
HBC 351                 & 0.61           & 8        & 85           & V892 Tau                & 4.0            & 13       & 560          \\
HBC 352/3               & 8.6            & 8        & 1200         & V927 Tau                & 0.29           & 8        & 41           \\
HBC 354/5               & 6.3            & 11       & 880          & V928 Tau                & 0.18           & 8        & 25           \\
HBC 356/7               & 2.0            & 11       & 280          & V955 Tau                & 0.33           & 8        & 46           \\
HBC 360/1               & 7.2            & 8        & 1010         & VY Tau                  & 0.66           & 8        & 92           \\
HBC 412                 & 0.70           & 8        & 98           & ZZ Tau                  & 0.03           & 12       & 4            
\enddata
\tablenotetext{a}{The table columns are: (1) - object name; (2) - projected 
separation in arcseconds for each pair (sb denotes a spectroscopic binary); 
(3) - reference for each separation measurement (see $b$); (4) - projected 
semimajor axis in AU, assuming a distance of 140\,pc (spectroscopic binary 
separations are the best-fit values of $a \sin i$).} 
\tablenotetext{b}{Projected separations, $\theta_p$, are average values from 
the following sources: 0 = \citet{mundt83}; 1 = \citet{rodriguez86}; 2 = 
\citet{simon87}; 3 = \citet{leinert89}; 4 = \citet{haas90}; 5 = 
\citet{leinert91}; 6 = \citet{simon92}; 7 = \citet{reipurth93}; 8 = 
\citet{leinert93}; 9 = \citet{ghez93}; 10 = \citet{richichi94}; 11 = 
\citet{mathieu94}; 12 = \citet{simon95}; 13 = \citet{leinert97}; 14 = 
\citet{mathieu97}; 15 = \citet{duchene99}; 16 = \citet{white01}; 17 = 
\citet{tamazian04}; 18 = \citet{duchene04}; 19 = \citet{martin04}; 
20 = \citet{white05}.}
\tablenotetext{c}{These high-order multiple systems were assigned various 
flux densities and disk masses for the analysis described in \S 3.5 and 
represented in Figure \ref{semimajor} and Table \ref{binary_prob}.  T Tau
--- The stellar components are too close to rule out a large circum-triple 
disk from current interferometric observations, so each separation is assigned 
the same flux/mass.  UX Tau --- High-resolution 1.3\,mm observations show that 
A is the only component with a disk \citep{jensen03}, so the 2 separations 
($a_p = 380$ and 830\,AU) are assigned the same flux/mass value, and the close 
binary B is given a flux density equal to the completeness limit (10\,mJy) and 
a corresponding disk mass from Equation 5.  UZ Tau --- We adopt the result of 
\citet{jensen96} that the W close binary contributes $\sim$19\% of the total 
flux density in the system, and scale the flux densities and disk masses 
accordingly for each separation.  HV Tau --- We assume that C is the sole 
component with a disk because the AB close binary has a Class III SED, and C 
has been shown in the optical to have an edge-on disk \citep{stapelfeldt03}.  
The AB separation ($a_p = 4$\,AU) is assigned a flux density equal to the 
completeness limit (10\,mJy) and a corresponding disk mass from Equation 5.}
\tablenotetext{d}{GG Tau: a quadruple source, these numbers refer only to the 
A close binary, as the B system falls outside the 850\,$\mu$m SCUBA beam.  GN 
Tau: \citet{white01} note a significantly larger projected separation, 
$\theta_p = 0\farcs33$.} 
\end{deluxetable}

\begin{deluxetable}{lcrr}
\tablecolumns{3}
\tablewidth{0pc}
\tabletypesize{\scriptsize}
\tablecaption{Multiplicity Effects on Disk Properties\tablenotemark{a} \label{binary_prob}}
\tablehead{
\colhead{samples} & \colhead{$a_c$\,[AU]} & \colhead{$P(F_{\nu})$} & \colhead{$P(M_d)$}}
\startdata
close vs. wide             & 50  & $85 - 91$\% & $33 - 54$\%   \\
close vs. single           & 50  & $ 9 - 60$\% & $\le 7$\%     \\
wide vs. single            & 50  & $53 - 76$\% & $53 - 80$\%   \\
close vs. wide             & 100 & $97 - 99$\% & $\ge 99$\%    \\
close vs. single           & 100 & $26 - 70$\% & $34 - 83$\%   \\
wide vs. single            & 100 & $77 - 91$\% & $65 - 89$\%   \\
\hline
Class II, close vs. wide   & 50  & $90 - 94$\% & $42 - 73$\%   \\
Class II, close vs. single & 50  & $75 - 90$\% & $16 - 33$\%   \\
Class II, wide vs. single  & 50  & $4 - 14$\%  & $89 - 96$\%   \\
Class II, close vs. wide   & 100 & $94 - 98$\% & $\ge 98$\%    \\
Class II, close vs. single & 100 & $83 - 90$\% & $91 - 98$\%   \\
Class II, wide vs. single  & 100 & $12 - 38$\% & $18 - 38$\%   
\enddata
\tablenotetext{a}{Results of censored statistical tests to determine how 
stellar companions affect circumstellar disks.  See also a graphical comparison 
in Figure \ref{semimajor}.  Two-sample tests were performed on the categories 
listed in the left column and described in the text (\S 3.5): close binaries 
(with $a_p \le a_c$), wide binaries (with $a_p > a_c$), and single stars.  The 
probabilities that the 850\,$\mu$m flux densities ($P(F_{\nu})$) or disk masses 
($P(M_d)$) are drawn from \emph{different} parent populations are given in the 
last two columns.  The ranges in the probabilities are representative of the 
various statistical tests.}
\end{deluxetable}

\clearpage

\begin{deluxetable}{lcccccccc}
\tablecolumns{9}
\tabletypesize{\scriptsize}
\tablewidth{0pc}
\tablecaption{Summary of Submillimeter Properties\tablenotemark{a} \label{summ}}
\tablehead{
\colhead{sample} & \colhead{$N_f$} & \colhead{$f_{\rm{smm}}$} & \colhead{$N_{M_d}$} & \colhead{median $M_d$\,[M$_{\odot}$]} & \colhead{$\sigma(M_d)$\,[dex]} & \colhead{$N_{\alpha}$} & \colhead{median $\alpha$} & \colhead{$\sigma(\alpha)$}}
\startdata
Class I       & 16  & $1.00 \pm 0.25$ & 16 & $3 \times 10^{-2}$ & 0.59 & 16 & 2.51    & 0.54 \\
Flat-Spectrum & 9   & $1.00 \pm 0.33$ & 9  & $4 \times 10^{-3}$ & 0.53 & 7  & 2.02    & 0.70 \\
Class II      & 74  & $0.86 \pm 0.11$ & 64 & $3 \times 10^{-3}$ & 0.67 & 34 & 1.79    & 0.51 \\
Class III     & 54  & $0.07 \pm 0.04$ & 4  & $6 \times 10^{-4}$ & 0.26 & 0  & \nodata & \nodata \\
WTTS          & 61  & $0.15 \pm 0.05$ & 9  & $2 \times 10^{-3}$ & 0.65 & 5  & 1.73    & 0.27 \\
CTTS          & 74  & $0.91 \pm 0.11$ & 67 & $4 \times 10^{-3}$ & 0.72 & 40 & 2.02    & 0.62 \\
multiples     & 61  & $0.66 \pm 0.10$ & 40 & $2 \times 10^{-3}$ & 0.74 & 19 & 2.13    & 0.52 \\
singles       & 92  & $0.58 \pm 0.08$ & 53 & $1 \times 10^{-2}$ & 0.64 & 38 & 2.01    & 0.60 \\
\hline
total         & 153 & $0.61 \pm 0.06$ & 93 & $5 \times 10^{-3}$ & 0.71 & 57 & 2.01    & 0.57
\enddata
\tablenotetext{a}{The $N$ values are the total numbers of sources in each 
subsample.  $f_{\rm{smm}}$ is the fraction of those objects which were detected 
at a submillimeter wavelength along with the 1-$\sigma$ Poisson counting 
error.  Median values of disk masses and submillimeter continuum slopes are 
given, along with the standard deviations ($\sigma$) of those values.  The 
$\sigma(M_d)$ values are on a log scale.}
\end{deluxetable}

\begin{deluxetable}{lrrr}
\tablecolumns{4}
\tabletypesize{\scriptsize}
\tablewidth{0pc}
\tablecaption{Outer Disk Evolution: Survival Analysis Tests\tablenotemark{a} \label{evol_prob}}
\tablehead{
\colhead{samples} & \colhead{$P(F_{\nu})$} & \colhead{$P(M_d)$} & \colhead{$P(\alpha)$}}
\startdata
Class I vs. Flat-Spectrum   & $63 - 85$\%  & $>99.2$\%    & $74 - 87$\% \\
Class I vs. Class II        & $>99.96$\%   & $>99.9999$\% & $>99.7$\%   \\
Class I vs. Class III       & $>99.9999$\% & $>99.9999$\% & \nodata     \\
Flat-Spectrum vs. Class II  & $89 - 94$\%  & $60 - 83$\%  & $1 - 36$\%  \\
Flat-Spectrum vs. Class III & $>99.9999$\% & $>99.9999$\% & \nodata     \\
Class II vs. Class III      & $>99.9999$\% & $>99.9999$\% & \nodata     \\
WTTS vs. CTTS               & $>99.9999$\% & $>99.9999$\% & $26 - 65$\%   
\enddata
\tablenotetext{a}{The results of survival analysis tests as described in the 
text (\S 4).  The values of $P(F_{\nu})$, $P(M_d)$, and $P(\alpha)$ are the 
probabilities that the 850\,$\mu$m flux densities, disk masses, and 
submillimeter continuum slopes, respectively, of the two samples in the first 
column are drawn from \emph{different} parent populations.  The ranges of $P$ 
values are representative of the various statistical tests.  See Figure 
\ref{diskevol} for a graphical representation of this comparison.}
\end{deluxetable}

\end{document}